\documentclass[%
reprint,
superscriptaddress,
nofootinbib,
amsmath,amssymb,
aps,
prb,
]{revtex4-2}
\usepackage[english]{babel}

\pdfoutput=1
\usepackage{float}
\usepackage{xcolor}
\usepackage{graphicx}
\usepackage{dcolumn}
\usepackage{bm,lipsum}
\usepackage{siunitx}
\usepackage[colorlinks=true, linkcolor=black, citecolor=black, urlcolor=black]{hyperref}
\usepackage{multirow}
\usepackage{tikz}
\usepackage{cancel}
\usepackage{orcidlink}
\usepackage{bibunits}
\usepackage{placeins}

\defaultbibliography{references}
\defaultbibliographystyle{apsrev4-2-titles} 

\usepackage{graphicx}
\usepackage{booktabs}
\usepackage{tabularx}
\usepackage{xcolor,colortbl}
\usepackage[T1]{fontenc}

\newif\ifmanycite
\manycitefalse

\newcommand{\om}[1]{\tilde{\omega}_\mathrm{#1}}
\newcommand{\gam}[1]{\gamma_\mathrm{#1}}
\newcommand{\rabi}{\Omega_\mathrm{R}}
\newcommand{\Om}{\tilde \Omega}
\newcommand{\Omp}{\tilde \Omega_p}

\newcommand{\cav}{\mathrm{c}}
\newcommand{\omcav}{\tilde \omega_\cav} %cavity mode frequency

\newcommand{\pos}{\mathbf{r}}
\newcommand{\fH}{\mathbf{H}}
\newcommand{\fB}{\mathbf{B}}
\newcommand{\fE}{\mathbf{E}}
\newcommand{\fD}{\mathbf{D}}
\newcommand{\fP}{\mathbf{P}}

\newcommand{\ii}{\mathrm{i}}
\newcommand{\epsr}{\varepsilon_\mathrm{r}}
\newcommand{\po}{(\pos, \omega)}

\newcommand{\coupling}{\kappa}
\newcommand{\hgg}{\coupling^2}
\newcommand{\hggi}{\coupling_p^2}
\newcommand{\coeff}[1]{\Phi_{#1}}
\newcommand{\dotcoeff}[1]{\dot{\Phi}_{#1}}
\newcommand{\veccoeff}{\mathbf{\coeff{}}}
\newcommand{\veccoeffrse}{\boldsymbol{\psi}}

\newcommand{\mb}[1]{\mathbf{#1}}

\newcommand{\Eq}[1]{Equation~\eqref{#1}}

\newcommand{\carsten}[1]{#1}

\newcommand{\important}[1]{#1}

\def\RS/{RS}
\def\RSs/{RSs}
\def\RSsintro/{resonant states (RSs)}
\def\RSlong/{resonant state}
\def\RSslong/{resonant states}
\def\QNMslong/{quasi normal modes}

\newcommand{\papertitle}{Resonant states reveal strong light-matter coupling in nanophotonic cavities}
\newcommand{\paperauthors}{
\author{{Jan David Fischbach }\orcidlink{0009-0003-8765-8920}}
\affiliation{Institute of Nanotechnology, Karlsruhe Institute of Technology, Karlsruhe, Germany}

\author{Sergei Gladyshev}
\affiliation{University of Graz, and NAWI Graz, Graz, Austria}

\author{Adrià Canós Valero}
\affiliation{University of Graz, and NAWI Graz, Graz, Austria}
\affiliation{Riga Technical University, Institute of Telecommunications, Riga, Latvia}

\author{Markus Nyman}
\affiliation{Institute of Theoretical Solid State Physics, Karlsruhe Institute of Technology, Karlsruhe, Germany}

\author{Thomas Weiss}
\affiliation{University of Graz, and NAWI Graz, Graz, Austria}

\author{Carsten Rockstuhl}
\affiliation{Institute of Nanotechnology, Karlsruhe Institute of Technology, Karlsruhe, Germany}
\affiliation{Institute of Theoretical Solid State Physics, Karlsruhe Institute of Technology, Karlsruhe, Germany}
\affiliation{Center for Integrated Quantum Science and Technology (IQST), Karlsruhe Institute of Technology, Karlsruhe, Germany}
}

\begin{document}
\begin{bibunit}

\title{\papertitle}
\paperauthors

\begin{abstract}
%\newsincearxiv{
Photonic resonances enable control over light-matter interactions, but many key phenomena only emerge in the strong-coupling regime where light and matter excitations fully hybridize. To distinguish between weak and strong coupling, one conventionally studies real-frequency spectra of the hybrid system. However, these spectra only provide indirect estimates of the underlying resonant dynamics, as the resonances reside at complex frequencies. To overcome this contradiction, we demonstrate that photonic resonant states provide a framework for unambiguously distinguishing between weak and strong coupling. Upon tracing the resonant states through the complex plane while changing the resonator geometry, their trajectories undergo a qualitative change at the onset of strong coupling. Instead of passing each other in the complex frequency plane with only perturbative interactions, the resonant states swap positions. Assuming a single dominant photonic resonance, we derive an effective Hamiltonian that captures the interaction with multiple material resonances, including direct access to coupling rates from overlap-integrals. Our analysis reveals that, unlike most coupled-oscillator models commonly employed, hybridization not only introduces off-diagonal coupling but also shifts the bare eigenfrequency of the photonic mode. We apply our approach to planar and spherical silver resonators filled with a molecular material whose properties were extracted from quantum-chemical simulations.
\end{abstract}
\maketitle
 
\begin{figure}[t]
    \centering
    \includegraphics[width=1\linewidth]{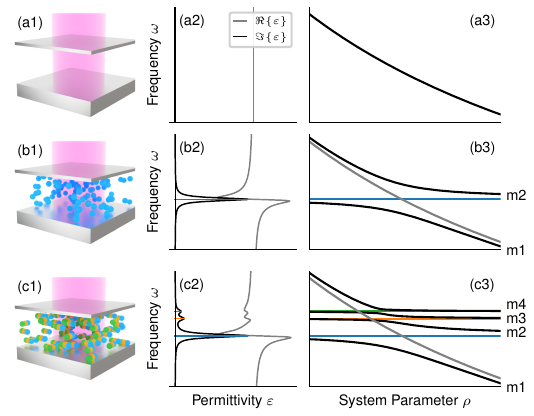}
    \caption{
    Mode hybridization: (a) When a system parameter $\rho$ of a photonic resonator is modified, the system dynamics change accordingly, expressed here in terms of a changing (complex) eigenfrequency as a function of $\rho$ \cite{LPR_Transient_Dynamics_2023}. Examples for possible choices of $\rho$ are the spacing between the plates for a Fabry-Perot cavity or the radius of a core or a shell for a spherical resonator. (b) When a material characterized by a single Lorentz-oscillator is introduced to the photonic resonator, the photonic mode, which can be considered as an independent oscillator, and material resonance couple to each other, leading to hybridization. The new hybrid modes distribute their energy across the coupled system, leading to shifts in the eigenfrequencies from the bare system without coupling. 
    %The splitting $\rabi$ is related to the coupling $\coupling$ by \Eq{eq:splitting}. 
    (c) The simple relation between coupling coefficients and frequency splitting gets spoiled when multiple material resonances are introduced. %characterize the introduced material.%, as is the case for almost all realistic photonic materials.
    Column (1) illustrates the considered system, column (2) expresses the material properties, and column (3) shows the (un-)coupled eigenfrequencies.
    % In these figures, the left column (1) illustrates the geometrical setting, and the color for the introduced material reflects the number of  resonances used to describe it. The color code will be used throughout the manuscript and shall ease orientation. The central column (2) expresses the properties of the material, the right column (3) the eigenfrequencies of the resonances. In the right column (3), the line color is linked to the different resonances (gray = optical mode; blue, red, and green = material resonaces), and black are the hybrid modes after hybridization. 
    }
    \label{fig:coupled_oscillators}
\end{figure}
\section{Introduction}
% \important{Strong coupling between material resonances and photonic resonances is one of the key effects in nanophotonics and quantum optics}
Hybrid states of light and matter form when photonic and material resonances are coupled [see Figure~\ref{fig:coupled_oscillators}]~\cite{weisbuch_observation_1992, lidzey_strong_1998, LPR_strong_coupling_2019}. When the coupling is strong, the photon and material resonances each contribute a significant part to the resulting hybrid modes, instead of just perturbing one another. These so-called polaritons possess emergent properties distinct from their constituents~\cite{aberra_guebrou_coherent_2012, LPR_nonlinear_strong_2021}. As such, they enable the development of advanced technologies, comprising efficient polariton-lasers~\ifmanycite\cite{butov_polariton_2007, schneider_electrically_2013, stoferle_ultracompact_2010, LPR_polariton_lasing_2021}\else\cite{butov_polariton_2007, schneider_electrically_2013}\fi, all-optical compute~\ifmanycite\cite{ballarini_all-optical_2013, tassan_integrated_2024, zasedatelev_room-temperature_2019, kavokin_polariton_2022}\else\cite{ballarini_all-optical_2013, zasedatelev_room-temperature_2019, kavokin_polariton_2022}\fi, optical neural networks~\ifmanycite\cite{ballarini_polaritonic_2020, mirek_neuromorphic_2021, opala_harnessing_2023}\else\cite{mirek_neuromorphic_2021}\fi, and high-resolution (label-free) sensors~\ifmanycite\cite{perez-gonzalez_optical_2013, xu_surface_2019, thomas_all-optical_2022, zheng_quantum_2023}\else\cite{thomas_all-optical_2022, baranov_toward_2023}\fi. Overall, the range of possible applications leveraging polaritons is vast, spanning from solar cells and LEDs~\cite{fei_controlling_2024} to platforms for topological physics~\ifmanycite\cite{liu_generation_2020, kavokin_polariton_2022}\else\cite{heimig_topological_2025, kavokin_polariton_2022}\fi. Recent advances in polaritonic chemistry have even shown that strong coupling can be leveraged to modify the reaction energy landscape to favor selected reaction pathways~\ifmanycite\cite{hutchison_modifying_2012, thomas_ground-state_2016, thomas_tilting_2019, hertzog_strong_2019, nagarajan_chemistry_2021}\else\cite{hutchison_modifying_2012, 
hertzog_strong_2019, baranov_toward_2023}\fi. 

Typically, these applications benefit from an increased coupling strength between photonic and material resonances. 

To assess the strength of light-matter coupling in optical systems, a standard approach is to compare the spectra of the uncoupled components -- the photonic resonator and the material -- with that of the coupled system. In the strong coupling regime, spectrally aligned bare-system resonances typically split into two spectrally separated peaks in the coupled spectrum. The spectral separation between these split peaks is widely used as a measure of the coupling strength.
However, accurately assessing the coupling strength is subtle, as the inherent non-Hermitian nature of open photonic systems, that can be probed from outside, blurs the view of the resonances~\cite{savona_quantum_1995, carlson_strong_2021, zerulla_multi-scale_2024, limonov_fano_2017}. 
For instance, the authors of Reference~\cite{savona_quantum_1995} showed how the splitting extracted from peaks/dips in the spectra of different observable quantities of the same system can differ widely. 
Moreover, many realistic materials feature multiple closely spaced material resonances~\ifmanycite\cite{kena-cohen_strong_2008, xiang_intermolecular_2020, zerulla_multi-scale_2024}\else\cite{zerulla_multi-scale_2024, heimig_topological_2025}\fi [see Figure~\ref{fig:coupled_oscillators}(c)], adding further complications to the extraction of coupling parameters from observable spectra. The fundamental problem is that when judging the coupling strength based on an optical quantity defined at a real frequency, we only indirectly probe the features of the resonances. The resonances in most open systems live at complex frequencies, and ultimately, the question of whether strong coupling emerges or not requires a consideration of the resonance interaction at these complex frequencies.

In this work, we distinguish weak and strong light-matter coupling at complex frequencies. Our analysis is based on the \RSsintro/ -- also known as \QNMslong/, decaying states, natural modes, etc. -- constituting a natural framework to describe photonic resonances. Unlike probing the optical response under some illumination at real frequencies, the study of the \RSs/ allows an explicit determination of the resonance frequencies and the damping rates of the hybrid structure, without the need for any fitting to optical spectra.

%\newsincearxiv{
When increasing the size of the photonic resonator, the \RS/-eigenfrequencies move through the complex plane. They emerge from singularities of the permittivity or at infinite frequency for a vanishingly small resonator and approach zeros of the permittivity or zero frequency for increasingly large size. A qualitative change in these trajectories marks the onset of what we call hidden strong coupling: In the case of weak coupling, the trajectories start and end in poles and zeros of the permittivity for modes mostly sustained by the material or in poles and zeros of the frequency for modes dominantly associated with the photonic mode. In the (hidden) strong coupling regime, on the other hand, the trajectory starting at infinite frequency ends up approaching the permittivity zero, swapping places with a trajectory starting at a permittivity pole that ends approaching zero frequency. This transition provides a binary signature to discriminate between the two regimes. Radiative and dissipative losses blur the view of these trajectories. These losses compete with the energy exchange between light and matter, impeding many of the sought-after effects. The fact that the split modes cannot be clearly discerned in measurable observables motivates denominating this regime as \emph{hidden} strong coupling, with the transition to \emph{observable} strong coupling occurring as the coupling starts to dominate over all the losses.
%}

In contrast to previous studies~\cite{christ2003a, canales_polaritonic_nodate, carlson_strong_2021, muljarov_resonant-state_2016, sehmi_applying_2020}, we derive \important{for the first time} an effective Hamiltonian that explicitly shows the degree of mixing between the material resonances and a single \RS/ of the bare photonic resonator \important{directly from macroscopic Maxwell's equations}, yielding a closed-form expression for the coupling rate. Crucially, we discover that, unlike conventional coupled-oscillator models, light-matter coupling also introduces an additional spectral shift of the bare photonic eigenfrequency.

We will start our discussion with a brief review of the theoretical background on Maxwell's equations, \RSslong/, pole-based permittivity models, and the coupled oscillator model. We will briefly discuss various criteria considered to classify whether a system exhibits strong coupling. A commonality between the criteria is the necessity to accurately evaluate the coupling rate. We go on to provide a \RS/ formulation of the coupling between a single \RS/ and multiple material resonances. 

We will then proceed to investigate example systems of increasing complexity. First, we will treat a planar cavity filled with a medium characterized by a single Lorentz-resonance. It serves to discuss how signatures of strong coupling are obscured by radiative and dissipative damping. Using the same example, we will show how \RSs/ allow us to discriminate between weak and strong coupling, even when optical observables yield contradictory conclusions. In our second example, we analyze a core-shell nanosphere, showing the validity of our approach for single nanoparticles. \important{For the first time, our method enables direct access to the individual coupling rates without the need for phenomenological coupled-oscillator models.} We go on to demonstrate how these rates are particularly useful when investigating systems that host multiple adjacent material resonances. In such systems, the overall splitting of the hybrid modes results from complex contributions of the different material resonances. With our method, these contributions can be disentangled from each other.

Our findings establish \RSs/ as a clear and universal framework for light-matter coupling in photonic resonators, avoiding ambiguous spectral criteria and paving the way to systematically design and engineer strong coupling in complex photonic systems.

\section{Results}
\subsection{Theoretical background}
\label{sec:theory}

% This contribution applies the rotating wave approximation. Further we assume only a single bare optical mode to contribute significantly. Under these assumptions a large class of photonic systems can be faithfully described. Nonetheless, in a later publication \cite{part2} we aim to generalize the presented derivation to an arbitrary number of initial modes including counterrotating contributions. 

% A detailed derivation of the relation between the QNMs of a system before and after introducing material resonances, under the given assumptions via the resonant state expansion (RSE) is provided in the appendix. However, starting from a more easily accessible coupled oscillator model an almost equivalent eigenvalue problem can be recovered:
%The RSE based derivation  minor correction will then be introduced

The propagation of electromagnetic waves in the frequency domain is governed by macroscopic Maxwell's curl equations~\cite{stamatopoulou_strong_2024}:
\begin{equation}
    \begin{aligned}
        \nabla \times \fH \po &= -\ii \omega \fD \po\, ,\\
        \nabla \times \fE \po &= \ii \omega \fB \po\, ,
    \end{aligned}
\end{equation}
where $\fE \po$ and $\fB \po$ are the electric and magnetic fields, respectively, while $\fD \po$ and $\fH \po$ are auxiliary fields. For readability, we will assume isotropic non-magnetic materials, \textit{i.e.},
\begin{equation}
\begin{aligned}
    \fB \po &= \mu_0 \fH \po\, ,\\
    \fD \po &= \varepsilon_0\fE \po + \fP \po\, ,
\end{aligned}
\end{equation}
where the polarization $\fP \po = \varepsilon_0 \chi \po \fE \po$ in the material is linearly related to the electric field by the susceptibility $\chi \po$. $\varepsilon_0$ and $\mu_0$ are the vacuum permittivity and permeability, respectively.

Describing resonant light-matter interaction naturally calls for the introduction of \RSs/ as the mathematical manifestation of photonic resonances. These \RSs/ are defined as solutions of the source-free Maxwell equations with radiation boundary conditions. Strictly speaking, \RSs/ also comprise optical resonances that are decoupled from the radiation continuum, known as bound states in the continuum~\cite{Hsu2016a, LPR_BIC_strong_2021}. However, these \RSs/ cannot be accessed by far-field measurements. For all other \RS/, the openness of the system necessarily leads to damping over time. In contrast to eigenmodes of closed systems or bound states in the continuum, the eigenfrequencies of \RSs/ are thus complex-valued, with the imaginary part determining the exponential damping rate in time. As a consequence, the eigenfields diverge away from the resonator, which results in several challenges, including their normalization, orthogonality, and completeness~\cite{both_resonant_2022, sauvan_normalization_2022, kristensen_modeling_2020}. If an observable existed that was coupled solely to a single \RS/ with eigenfrequency $\omcav = \omega_\cav-\ii \gamma_\cav$, the observable would exhibit a Lorentzian lineshape centered around the real part of that eigenfrequency (we call $\omega_\cav$ its resonance frequency) and broadened according to the imaginary part of its eigenfrequency (we call $\gamma_\cav$ its damping rate). However, \RSs/ can interfere with each other and with a non-resonant background, forming intricate Fano lineshapes \cite{lukyanchuk_fano_2010, miroshnichenko_fano_2010, sauvan_theory_2013, bai_efficient_2013, limonov_fano_2017}.

% Photonic cavities, that  are open systems and as such inherently non-Hermitian. The lack of hermiticity results in a number of challenges, when working with the damped eigenmodes, including the questions of normalization, orthogonality and completeness. A particular challenge is posed by the fact that damped eigenmodes have fields that grow exponentially in space, as a consequence of the damping in time. .  

% On the basis of such QNMs a theory to rigorously find the QNMs of a system subject to arbitrarily strong perturbations has been put forward. This so called resonant state expansion (RSE) faithfully reproduces the QNMs of the perturbed system under the condition that a sufficiently large number of initial QNMs is considered \cite{muljarov_brillouin-wigner_2011}. 

% \subsection{Material Models}
Now that we have laid the foundation for the treatment of electromagnetic resonances, let us switch our attention to the description of resonant material excitations. While a rigorous \textit{ab-initio} description requires solving the interaction of light and matter quantum-mechanically~\cite{koenderink_single-photon_2017}, semi-classical descriptions boil down to the movement of bound charges $q$ (with effective mass $m_j$; the index $j$ indicates different species) that are displaced from their equilibrium positions in the effective potential landscape of the inner electrons and the atomic nuclei by $\mathbf{x}(t)$ due to the presence of a driving field. Solving the corresponding differential equations yields the induced polarization $\fP(t) = N_j q_j \mathbf{x}(t)$ with number density $N_j$. In the frequency domain, this results in the well-established Lorentz model~\cite{fox_optical_2012}:
\begin{equation} \label{eq:lorentz_model}
\begin{aligned}
\epsr(\omega) &= 1+\chi(\omega)\\ &= 1 + \sum_j \frac{N_jq_j^2}{m_j \varepsilon_0} \frac{f_j}{\omega_{0,j}^2 - \omega^2 - \ii \Gamma_j\omega}\, .
\end{aligned}
\end{equation}
Here, $f_j$ is the strength of the $j^{\text{th}}$ oscillator with resonance frequency $\omega_{0,j}$ and damping $\Gamma_j$. Additional Drude-terms can be added to account for unbound charges~\cite{hecht_optics_2012}.  

 To unravel the contribution of each material resonance, we write Equation~(\ref{eq:lorentz_model}) as a sum of resonant contributions \cite{ben_soltane_generalized_2024}
\begin{equation} \label{eq:material_poles}
\epsr(\omega)= \varepsilon_\infty - \sum_p \frac{\varepsilon_p}{\omega - \Om_p}\, ,
\end{equation}
where $\varepsilon_p$ corresponds to the excitation strength of the $p^{\text{th}}$ material resonance, $\Om_p$ is the complex frequency of the material-pole, and $\varepsilon_\infty$ encapsulates the background permittivity resulting from resonances far outside the frequency range of interest. For convenience, we provide the expressions for $\varepsilon_p$ and $\Om_p$ linking Equation~\eqref{eq:material_poles} to Equation~\eqref{eq:lorentz_model} in the Supplementary Information (SI).

\subsection{Criteria for strong coupling in open systems}
\label{sec:coupled_oscillators}

% In Section \ref{sec:resonant_states} and the Supplementary Information we provide a rigorous derivation of the interaction between a mode of an optical cavity (also known as resonator) and one or more of such material resonances.
 
 To gain first insights into the physics of light-matter coupling, we will first consider a standard phenomenological model of two coupled harmonic oscillators. As open systems are inherently lossy, we will include a damping in the form of complex-valued initial (bare) eigenfrequencies of the optical cavity $\omcav = \omega_\cav-\ii \gamma_\cav$ and the material resonance: $\Om_\mathrm{p}=\omega_\mathrm{p}-\ii \gam{p}$. The eigenfrequency of the cavity mode $\omcav(\rho)$ is assumed to be a function of a system parameter $\rho$ -- such as the cavity thickness in a Fabry-Perot (FP) cavity. The coupled equations of motion describe the joint dynamics due to the light-matter coupling $\kappa$ \cite{carlson_strong_2021}:
\begin{equation}
\label{eq:small_hamiltonian}
\omega
\underbrace{\begin{pmatrix}
\coeff{\cav}(\rho)\\
\coeff{\mathrm{p}}(\rho)
\end{pmatrix}}_{\veccoeff(\rho)}= 
\underbrace{
\begin{bmatrix}
\omcav(\rho) & \kappa(\rho) \\
\kappa(\rho) & \Om_\mathrm{p}
\end{bmatrix}}_{\mathcal{H}}
\begin{pmatrix}
\coeff{\cav}(\rho) \\
\coeff{\mathrm{p}}(\rho)
\end{pmatrix}\, .
\end{equation}
Here, $\coeff{\cav}(\rho)$ and $\coeff{\mathrm{p}}(\rho)$ are coefficients quantifying the contributions of the cavity mode and material resonance to the hybrid system~\cite{hertzog_strong_2019}. For readability, the $\rho$ dependence will not be stated explicitly throughout the rest of the derivation. By diagonalizing $\mathcal{H}$, we obtain uncoupled equations of motion for superpositions of the initial modes (considering the material resonance as a further mode of the system). These quasi-particles are commonly referred to as polaritons. They evolve with distinct eigenfrequencies
\[
\om{1,2} = \frac{\Om_\mathrm{p} + \omcav}{2} \pm \frac{1}{2}\sqrt{4\hgg + (\Om_\mathrm{p} - \omcav)^2 }\, .
\]
On resonance ($\omega_\cav=\omega_\mathrm{p}=\omega_0$), we get
\begin{equation}
\begin{aligned}
&\om{1,2} = \omega_0 - \frac{\ii(\gam p + \gamma_\cav)}{2} \pm \frac{\rabi}{2}\, ,\\
&\text{with } \rabi := \sqrt{4\hgg-(\gam p -\gamma_\cav)^2}\, .
\end{aligned}
\label{eq:splitting}
\end{equation}
For real-valued $\hgg$, \textit{i.e.}, retardation-free coupling, and
\begin{equation}
    \hgg > \frac{1}{4}(\gam p - \gamma_\cav)^2,
    \label{eq:strong_coupling}
\end{equation}
the splitting $\rabi$ is real-valued. Here, we consider \Eq{eq:strong_coupling} to mark the transition from \textbf{weak} to \textbf{(hidden) strong coupling}.
%(as do \cite{schneider_two-dimensional_2018}).
Damping-induced linewidth broadening obscures the splitting of resonance frequencies to a point that it may not be observed in typical experimental measurements, leading us to introduce the term ``hidden.'' Note how at the same time, the existence of damping is a prerequisite for the existence of the weak-coupling regime. To compensate for the effects of damping, a collection of alternative criteria for strong coupling has been used throughout the literature~\cite{zheng_manipulating_2017, wen_room-temperature_2017, rider_something_2021, stamatopoulou_strong_2024, torma_strong_2014, lalanne_light_2018, rider_strong_2024, novotny_principles_2012, barnes_strong_2023, menghrajani_molecular_2024, schneider_two-dimensional_2018, hertzog_strong_2019}, which will be further discussed in a dedicated section. $\rabi$ is commonly denoted as Rabi frequency, analogous to Rabi oscillations in optically-excited atoms. It can be understood as the frequency at which energy is exchanged between the coupled modes. Figure~\ref{fig:coupled_oscillators}(b) illustrates the resulting two-branched dispersion, featuring an avoided crossing with a separation corresponding to $\rabi$. Notice how the spectral separation increases with detuning sufficiently far from resonance, when $|\Om_\mathrm{p}-\omcav (\rho)|$ dominates over $\coupling$ in the discriminant. In experimental observations, where only the dynamics of the coupled system are accessible, it is thus common to assume that the resonance frequency and the frequency of minimum separation between the upper and lower polariton branches (UP/LP) coincide. We remark that in cases where the coupling parameter $\coupling(\rho)$ and/or cavity loss $\gamma_\cav (\rho)$ change significantly with $\rho$, the minimum separation between UP and LP is shifted away from the configuration $\rho$ for which the real parts of the bare eigenfrequencies coincide~\cite{symonds_particularities_2008}.

\subsection{Observable versus ``hidden'' strong coupling}\label{sec:strong_coupling_criteria}

The above treatment already considered loss channels originating from the photonic resonator and dissipation in the material. However, their implications on observable physics remain to be clarified. Effects relying on the coherent exchange of energy between modes are suppressed when the decay dominates. Applications in quantum information are particularly vulnerable to the resulting infidelity \cite{ghosh_quantum_2020}. In classical experiments, the spectral position of the resonant modes is most often determined from peaks or dips in spectrally-resolved observables such as reflection and transmission, or derived quantities such as absorption and, in the case of a few microscopic (plasmonic) resonators, \mbox{scattering-,} \mbox{extinction-,} and absorption-cross-sections. In these measurements, the predominant fingerprint of considerable loss $\gamma_{\mathrm{p},\cav}$ is a broadening of the observed line shape. As a consequence, the avoided crossing of coupled \RSs/ that fulfill the strong-coupling criterion [\Eq{eq:strong_coupling}] can be experimentally indiscernible.
This motivates the introduction of an additional regime, when the coupling rate dominates over all present loss channels. 
Henceforth, we consider a system with one material resonance to exhibit \textbf{observable strong coupling} if, in addition to \Eq{eq:strong_coupling}, the hybrid modes are sufficiently separated to be discernible despite their broadening.

The discernibility of adjacent broadened peaks is a lively field of ongoing scientific activity \cite{robertson_quantifying_2013, schulze_critical_2022, hagiwara_precision_2025}. As a result, the criterion for observable strong coupling is less clearly defined than the transition from weak to strong coupling.  For practicality, we deem superimposed Lorentzian lineshapes discernible if their centers are separated further than the sum of their intensity half widths at half maximum ($\gamma$), which translates to the criterion  (considering the factor $\frac{1}{2}$ in the definitions of $\gamma$, see \cite{zheng_manipulating_2017, wen_room-temperature_2017, rider_something_2021, stamatopoulou_strong_2024, torma_strong_2014}): 
\begin{equation}    
\frac{\rabi}{2} > \frac{\gam p + \gamma_\cav}{2}\, .
\label{eq:observable_strong_coupling_rabi}
\end{equation}
Using the definition in \Eq{eq:splitting}, it is possible to rewrite this criterion in terms of the coupling $\hgg$ \cite{rider_something_2021, stamatopoulou_strong_2024}:
\begin{equation}    
\hgg > \frac{\gam p^2 + \gamma_\cav^2}{2}\, .
\label{eq:observable_strong_coupling_g}
\end{equation}

% It is relatively common in literature to consider the criterion equivalently expressed in Eqs.~\eqref{eq:observable_strong_coupling_rabi} and \eqref{eq:observable_strong_coupling_g} as the transition from weak to strong coupling \cite{lalanne_light_2018, rider_strong_2024, torma_strong_2014}\footnote{Please note how it is frequently assumed that $2\coupling = \rabi$, which only holds under the condition that $\coupling \gg |\gam p - \gamma_\cav|$.}. Some authors consider additional criteria e.g. on the cavity finesse \cite{menghrajani_molecular_2024}. Which criterion should be used depends on the intended application/phenomena to be studied \cite{torma_strong_2014}. For example, when a coherent exchange of energy between light and matter is required the coupling rates $g$ can be required to exceed \textbf{any} of the damping rates $\gamma_\cav, \gam p$ of the bare system \cite{rider_something_2021, barnes_strong_2023} (in the following denoted as \textbf{energy-exchange strong coupling}). Moreover, the boundary between weak and strong coupling regimes is sometimes not considered to be sharp \cite{novotny_principles_2012}, reflecting the fact that observable spectra change continuously and not abruptly at the transition. 

Unfortunately, optical measurements (such as transmission or reflection measurements) do not provide direct access to the damping rates and the coupling coefficients that enter Equation~(\ref{eq:strong_coupling}) and Equation~(\ref{eq:observable_strong_coupling_g}). This is especially true for rich spectra with more than two resonances in the same spectral region, making interpretation challenging regardless of whether the coupling is weak, strong, or observably strong. As discussed next, there is even an ambiguity in the extraction of $\rabi$ from observable spectra.  Nonetheless, it is the established practice to find these parameters from fitting effective coupled-oscillator models to experimental or simulated spectra~\cite{zheng_manipulating_2017, wen_room-temperature_2017, stamatopoulou_strong_2024}.

\begin{figure*}[ht!]
    \centering
    \begin{tikzpicture}[      
        every node/.style={anchor=south west,inner sep=0pt},
        x=1mm, y=1mm,
      ]   
     \node (fig1) at (0,0)
       { \includegraphics[width=1\linewidth]{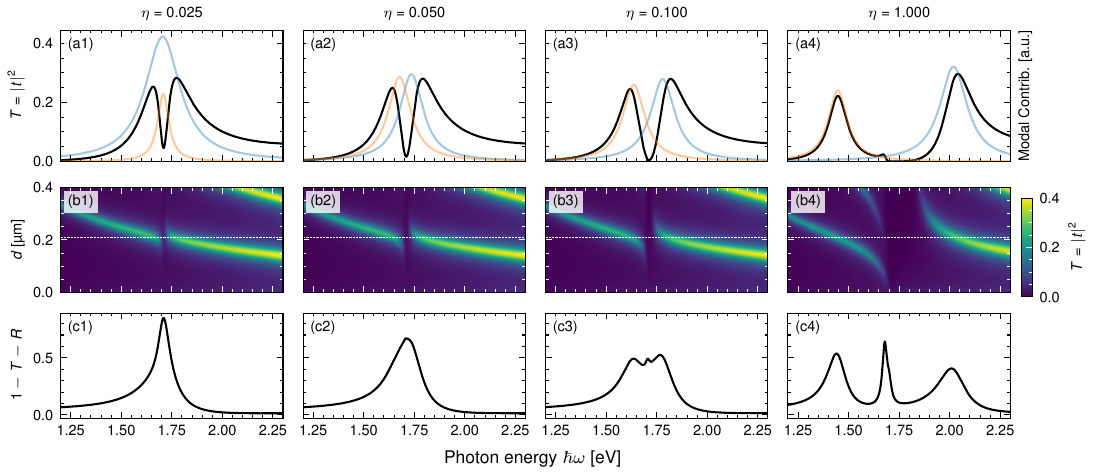}};
     \node (fig2) at (155,02) %(25,60)
       {\includegraphics[width=0.18\linewidth]{fig2_cavity_2_d.png}};  
    \end{tikzpicture}
    \caption{Optical response of a planar cavity filled with a resonant medium in terms of observables commonly used as hallmarks of strong coupling. From left to right (1-4), the coupling strength is varied by scaling the oscillator strength of an artificial single-pole Lorentz medium by a factor $\eta$ (parameters in Table~\ref{tab:params}). (a) Transmission through a cavity of thickness $d_\mathrm{res}\approx\qty{209}{\nano \meter}$ upon illumination from below. The orange and blue lines illustrate the contributions of the two dominant \RSs/, plotted as $\propto\left| \omega - \om{m}\right|^{-2}$ (the Lorentzian line shapes the \RSs/ would have in isolation). (b) The cavity mode is tuned by changing the cavity thickness $d$ ($d_\mathrm{res}$ indicated by dashed white line). % Note how (b1) resembles an avoided crossing, and (a1) correspondingly shows two clearly separated peaks, despite the system being in the weak coupling regime. The perceived splitting results from the interference between two modes that are at the same real frequency but have different linewidths. This phenomenon is known in the framework of Fano resonances~\cite{limonov_fano_2017}. The broad cavity mode (increasing transmission) can be understood as the background, on which a narrow absorption dip arises due to the material resonance. 
    (c) Absorption upon the same illumination and for the same cavity thickness $d_\mathrm{res}$ as in (a).}
    \label{fig:observable}
\end{figure*}

\begin{figure*}[t]
    \centering
    \begin{tikzpicture}[      
        every node/.style={anchor=south west,inner sep=0pt},
        x=1mm, y=1mm,
      ]   
     \node (fig1) at (0,0)
       { \includegraphics[width=1\linewidth]{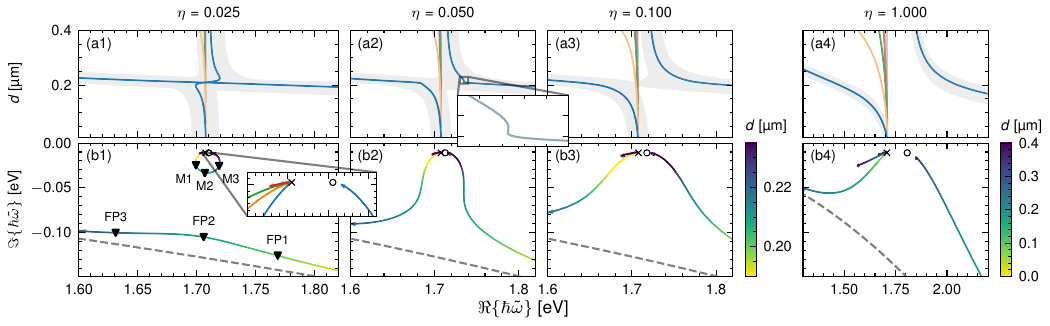}};
     \node (fig2) at (160,33)
       {\includegraphics[width=0.16\linewidth]{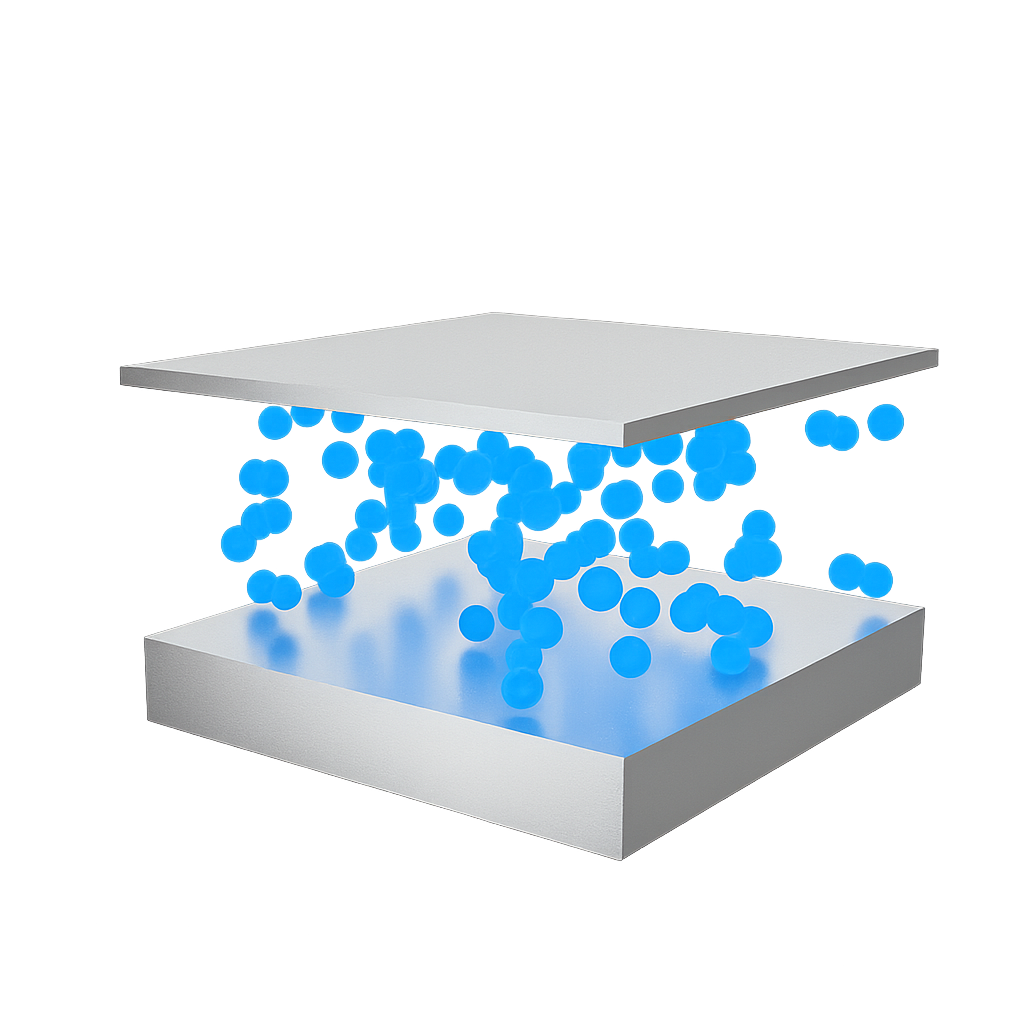}};  
    \end{tikzpicture}
    \caption{
    Parametrized pole trajectories for a planar cavity containing a single-pole Lorentz medium. As in Fig.~\ref{fig:observable}, panels 1-4 correspond to increasing oscillator strength (scaled by $\eta$).
    (a) Real parts of the \RS/ eigenfrequencies versus cavity thickness $d$ (different colors: cavity modes of different orders; blue being the fundamental mode). Each panel contains two branches, corresponding to distinct \RSs/ resulting from the hybridization. Gray shading shows $\Re\{\hbar \tilde \omega\} \pm \Im \{\hbar \tilde \omega\}$ of the fundamental mode.
    (b) Trajectories of the eigenfrequencies in the complex frequency plane as $d$ is varied (color encodes $d$). The dashed line indicates the dispersion of the bare cavity mode. A black cross (x) indicates the position of the single resonance of the Lorentz permittivity. A circle (o) marks its complex zero. The inset in (b1) shows additional details close to the permittivity pole making the higher-order \RSs/ more clearly visible (same color coding as in the top row). The triangle markers in (b1) are used in the main text to refer to sections of the trajectories when explaining these phenomena. FP and M denote hybrid modes that are dominated by Fabry-Perot and material contributions, respectively. The peculiar feature in the upper polariton branch in subfigure (b2; magnified in inset) results from the vicinity to the transition from weak to strong coupling, as it causes the eigenfrequency to locally vary almost exclusively in its imaginary part with changing $d$.
    }
    \label{fig:osc_reduction}
\end{figure*}

\subsection{From weak to observable strong coupling}
\label{sec:weak_strong_coupling}

To motivate the need to use \RSs/ in the investigation of light-matter coupling, let us start by analyzing a simplified system featuring a single material pole coupled to a Fabry-Perot mode of a planar cavity. The details of the system are summarized in Table~\ref{tab:params}. For illustrative purposes, we will operate the system in the three coupling regimes and traverse from \emph{weak coupling} to \emph{hidden strong coupling}, and finally to \emph{observable strong coupling}. To achieve such a transition between the different coupling regimes, we artificially reduce the oscillator strength $f_j$ or equivalently the residue of the corresponding material pole $- \varepsilon_\mathrm{p}$, and, with it, decrease the coupling between light and matter by multiplying the oscillator strength by a factor $\eta$ of 0.025, 0.05, 0.1, and 1, respectively. 

In an attempt to study the coupling regime, one would typically collect spectra of some observable quantity. We illustrate this approach by considering scattering-matrix-based thin-film calculations of the cavity under normal incidence from the side with the thinner silver layer. In Figure~\ref{fig:observable}(a1-a4), we show the transmission $T$ in black. Two peaks separated by a valley are present for all $\eta$. With increasing coupling strength, the frequency separation between the peaks increases. We want to emphasize that the presence of such a splitting does not prove the existence of strong coupling, as previously noted, \textit{e.g.}, in Reference~\cite{hertzog_strong_2019}. In particular, the splitting observed in Figure~\ref{fig:observable}(a1) results from Fano-type modal interference. It is readily understood from a basic model~\cite{limonov_fano_2017}: The narrow molecular absorption line carves a dip into the broad transmission peak of the Fabry-Perot mode, leaving the impression of a splitting. 

To shed light on the origin of this behavior, we foreshadow the \RS/-based analysis introduced later. The orange and blue lines in Figure~\ref{fig:observable}(a1-a4) show the two dominant \RS/ contributions to the transmission. The hybrid modes share the same center frequency and closely resemble the bare material resonance (orange) and the broader bare FP mode (blue). This gives us a first indication that for $\eta=0.025$ (column 1 of Figure~\ref{fig:observable}) the light-matter coupling is in the weak coupling regime. The perceived splitting results from the interference between these two \RSs/.

With increasing coupling strength, the observed peaks increasingly correspond to individual \RSs/, which becomes clear considering the good spectral alignment between the contributions from the individual \RSs/ with the full optical response in Figure~\ref{fig:observable}(a4). The only indication of the broken correspondence between peaks and individual \RSs/ at low coupling strengths is the non-Lorentzian lineshapes of the transmission [see Figure~\ref{fig:observable}(a1-a3)]. 

As a remedy, the authors of Reference~\cite{hertzog_strong_2019} suggest measuring the spectrum as a function of some changing system parameter and observing the avoided crossing of the polariton branches (as schematically illustrated in Figure~\ref{fig:coupled_oscillators} for a varying system parameter $\rho$) \cite{berkhout_simple_2020, cao_strong_2021}. In the current example of a Fabry-Perot cavity, we vary the cavity thickness $d$. The $d$-resolved transmission is shown in Figures~\ref{fig:observable}(b1-b4). Figures~\ref{fig:observable}(b3-b4) clearly display an avoided crossing, which is consistent with our later analysis, classifying these cases as \emph{observable strong coupling}. However, Figures~\ref{fig:observable}(b1) and (b2) also resemble an avoided crossing, which turns out to be a \emph{weak coupling} and a \emph{hidden strong coupling}, respectively. Different observable quantities, such as the Purcell factor, photoluminescence, or derived quantities, such as the absorption, depicted in Figures~\ref{fig:observable}(c1-c4), will exhibit peaks at different energies \cite{savona_quantum_1995, schneider_two-dimensional_2018, carlson_strong_2021}. Such an ambiguity necessitates a unified way of assessing the coupling between light and matter. In the following, we will take such an approach using the \RSs/, which are intrinsic to the resonator -- they do not depend on the external excitation.

We performed the same analysis as in Figure~\ref{fig:observable}(b), but tracked the evolution of the \RSs/ instead of the observable spectra (see also \cite{canales_polaritonic_nodate, ryabkovPolaritonicSpectraOptical2026}). \important{The scattering matrix based solver for layered media allows us to evaluate the scattering matrix at complex frequencies. We use a selective domain subdivision scheme \cite{bruno_evaluation_2024} with the AAA algorithm \cite{nakatsukasa_aaa_2018,betz_efficient_2024} and iterative sample refinement \cite{fischbach_framework_2025} to locate the poles of the determinant of scattering matrix, corresponding to the \RSs/}. Figures~\ref{fig:osc_reduction}(a1-a4) show how the real part of the eigenfrequencies (\textit{i.e.}, the resonance frequency) evolves with $d$. \RSs/ that originate from the same bare cavity mode (\textit{i.e.}, FP mode) are shown in matching colors (blue, orange, green, and red). Only for the fundamental FP mode, both hybrid modes (blue lines) fall within the plotted spectral window. Of the higher-order modes (orange, green and red), only one branch is present close to material resonance, while the other branch is far detuned. For now, we will focus our discussion on the \RSs/ related to the fundamental mode. In Figure~\ref{fig:osc_reduction}(a1) the highly dispersive mode consists predominantly of the fundamental FP mode, and the less dispersive mode is dominated by the material resonance. It is immediately clear that these two \RSs/ cross in Figure~\ref{fig:osc_reduction}(a1), while they avoid the crossing for (a2-a4), which corresponds to the transition from \emph{weak} to \emph{(hidden) strong coupling} according to \Eq{eq:strong_coupling}. For $\eta=0.025$ [Figure~\ref{fig:osc_reduction}(a1)], the \RSs/ cross because their difference in damping outweighs their strongly suppressed mutual coupling. 
%adiabatic transition does not lead to exchange of energy betweeen resonators. 

Figures~\ref{fig:osc_reduction}(b1-b4) introduce an additional perspective by tracing the trajectories of the \RSs/ through the complex frequency plane with changing $d$. As such, Figures~\ref{fig:osc_reduction}(a1-a4) can be understood as the projection of Figures~\ref{fig:osc_reduction}(b1-b4) onto the real frequency axis. The dispersion of the \RSs/ associated with the fundamental cavity mode (blue lines in Figure~\ref{fig:osc_reduction}(a)) is most prominent, as it resonantly couples to the material in the chosen thickness range. At the same time, the \RSs/ associated with higher-order cavity modes barely move, resulting in short line segments. These segments only have an appreciable length in Figure~\ref{fig:osc_reduction}(a4) due to the larger interval in which the cavity thickness $d$ is varied. In the remainder of this discussion will concentrate on the polariton branches emerging from the fundamental cavity mode.

The pole trajectories in Figure~\ref{fig:osc_reduction}(b1) reveal how the weakly-coupled FP mode modifies the material resonance. Together, they form hybrid states that are still clearly associated with either the FP mode or material resonance, respectively. This becomes apparent considering the asymptotic behavior when the cavity is far detuned to both extremes: The lower \RS/ approaches the resonance of the photonic cavity (dashed line) while the other approaches the material pole (black cross) or zero (black circle) for small and large cavity thickness, respectively. Henceforth, we refer to them as quasi-FP (qFP) and quasi-material (qM) modes, respectively. For small $d$, the qFP mode is blue-detuned from the material resonance [FP1; triangle marker in Figure~\ref{fig:osc_reduction}(b1)], while the coupling red-shifts the qM mode (akin to a Lamb shift; M1). When the real part of both modes matches (FP2; M2), predominantly the damping rate is modified (analogous to Purcell enhancement). After the qFP mode has passed by (FP3) upon increasing $d$ even further, the qM mode gets blue-shifted (Lamb shift; M3). 

\begin{figure*}[t]
    \centering
    \begin{tikzpicture}[      
        every node/.style={anchor=south west,inner sep=0pt},
        x=1mm, y=1mm,
      ]   
     \node (fig1) at (0,0)
       { \includegraphics[width=1\linewidth]{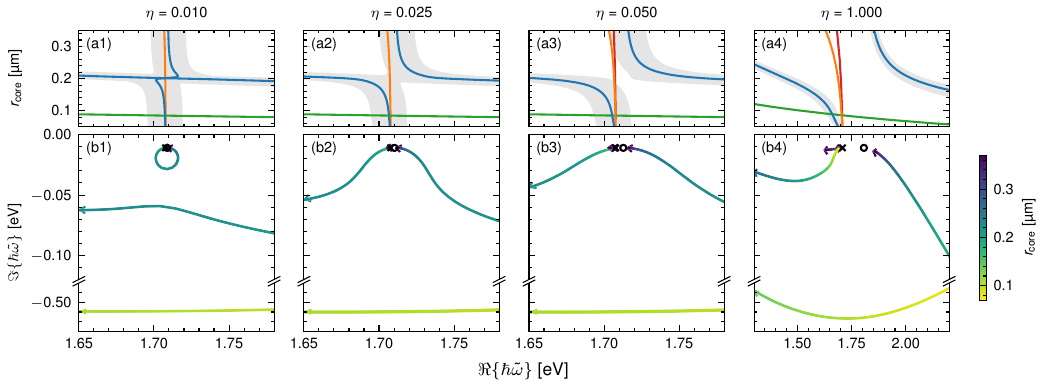}};
     \node (fig2) at (155,41) %(33,49)
       {\includegraphics[width=0.15\linewidth]{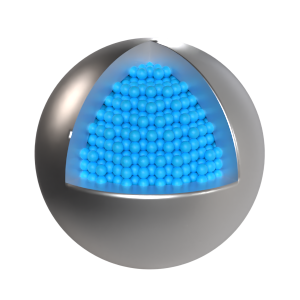}};  
    \end{tikzpicture}
    \caption{Parametrized pole trajectories for a 3D finite system. In analogy to Figure~\ref{fig:osc_reduction}, we trace the transverse magnetic dipolar resonances of a core-shell nanoparticle with a SURMOF core of variable radius $r_\mathrm{core}$ and a fixed 20~nm silver shell. As the light matter coupling is varied across the columns, the plasmon mode predominantly localized at the inner surface of the shell [blue/teal lines in (a)/(b) respectively] traverses from the \emph{weak} to the \emph{observable strong coupling} regime.}
    \label{fig:core_shell}
\end{figure*}

After the transition to \emph{(hidden) strong coupling} [Figure~\ref{fig:osc_reduction}~panel~2], the character of the hybrid modes is no longer clearly defined. Indeed, we see an entirely different behavior. 
The resonance frequency of one hybrid mode no longer crosses the other when sweeping the system parameter. Instead, we clearly observe that the hybrid modes smoothly exchange places in the complex frequency plane throughout the change of $d$. While the hybrid mode on the right (high energies) starts as a mostly FP-type mode for a small cavity thickness $d$, it ends up approaching the material zero (indicated by the black circle) for large $d$. At the same time, a \RS/ progressively emerges from the material pole (black cross) for a small thickness $d$ and approaches the trajectory of the bare FP mode for large $d$ (dashed line). Corresponding to Equation~\eqref{eq:strong_coupling}, the emergence of this distinct behavior is abrupt. Instead of just mutually perturbing one another, the hybrid modes exchange places, which constitutes an unambiguous signature of \emph{strong coupling}.
%the transition from the qFP mode passing by to the modes exchanging is abrupt. 
Exactly at the transition, the two \RSs/ coalesce, marking an \textbf{exceptional point}, where the square root in Equation~\eqref{eq:splitting} vanishes. Similarly, the authors of References~\cite{canos_valero_bianisotropic_2024, canales_polaritonic_nodate} observed the transition from weak to strong coupling of two geometrically coupled optical \RSs/ (far from any material resonance) through an exceptional point.  

At $\eta=0.05$ [Figure~\ref{fig:osc_reduction}(2)], the system is still quite close to the transition between \emph{weak} and \emph{hidden strong coupling}, which leads to a section of the trajectory, for which the resonance frequency -- \textit{i.e.} the real part of the complex eigenfrequency -- is almost unchanged [seen in the near-vertical movement of the line in Figure~\ref{fig:osc_reduction}(b2)]. In the projection onto the real axis [Figure~\ref{fig:osc_reduction}(b2)], this leads to a peculiar feature in the modal dispersion, as highlighted in the inset.

From Figure~\ref{fig:osc_reduction}(a), it becomes clear, however, that the loss-induced linewidth broadening obscures the emerging avoided crossing. The broadening is illustrated by the gray-shaded regions that span $\omega \pm \gamma$ for each \RS/, and that merge at the point of the avoided crossing. Equation~\eqref{eq:observable_strong_coupling_rabi} is fulfilled as soon as there is no overlap between the gray regions associated with the different polariton branches. For $\eta=0.1$ (panels 3 in Figures~\ref{fig:observable} and \ref{fig:osc_reduction}), this condition is on the verge of being met. Note how it coincides with the appearance of discernible peaks in the absorption spectrum [Figure~\ref{fig:observable}(c3)]. In addition, a third peak emerges and becomes more pronounced for $\eta=1$ (panels 4). The origin of this third peak is the subject of ongoing debate~\cite{antosiewicz_plasmonexciton_2014, stete_optical_2023, kondorskiy_optical_2024, greten_strong_2024} (also observed in Reference~\cite{tserkezis_mie_2018}). From the spectral analysis in Figure~\ref{fig:observable}(b4) and the modal dispersion of \RSs/ in Figures~\ref{fig:osc_reduction}(a4) and (b4), it becomes apparent that the central peak originates from higher-order cavity modes. In particular, the higher-order modes of the bare cavity are highly detuned from the material resonance, resulting in hybrid modes that closely resemble the bare modes. As a consequence, their material branches sit very close to the original material resonance, jointly creating the central peak. 
In the complex plane, this collection of higher-order optical modes corresponds to an \textbf{accumulation point} of \RSs/ at the material resonance, where the permittivity of the Lorentz model diverges. As a result, modes of arbitrarily high order (\textit{i.e.}, short wavelength in the medium) fit into the cavity. Such accumulation points are visible in Figures~\ref{fig:osc_reduction} to~\ref{fig:three_pole}.

% Finally, from the last column of Figure~\ref{fig:osc_reduction} it is obvious that the planar cavity filled with the simplified 1-pole SURMOF material clearly exhibits \emph{observable strong coupling}. 

% \subsubsection{Case of a single nanoparticle}
Before investigating systems with multiple material resonances, let us briefly shift attention to a system of finite size to highlight that the observed behavior is fundamental and not linked to a particular instance of a specific resonator. In this second example, we choose a spherical resonator --  specifically a core-shell nanoparticle (NP) \cite{tserkezis_mie_2018}. The core radius is chosen as the system parameter $\rho$ to tune the properties of the bare system. Material and geometry parameters are again given in Table~\ref{tab:params}. \important{To find the \RSs/ of the NP, we search for the eigenvalues of the characteristic matrix, by a combination of the contour integral method \cite{beyn_integral_2012} and an iterative refinement \cite{bykov_numerical_2013}}. The core of the NP is made of the same artificial material as in the previous example with optical properties that are characterized by a single pole. The shell is made of silver.

In analogy to Figure~\ref{fig:osc_reduction}, Figure~\ref{fig:core_shell} shows the trajectories of the \RS/-eigenfrequencies, in this case depending on the core radius. In the subfigures (a1-a4), the real part of the eigenfrequencies are shown as a function of the core radius, while in subfigures (b1-b4) the trajectories in the complex plane are shown (the color gradient of the line encodes the core radius). Because of the spherical symmetry, distinct multipoles and polarizations are orthogonal to each other and can thus be treated individually. To enhance the readability and clarity of presentation, we restrict ourselves to the transverse magnetic dipolar resonances.

The same qualitative change in the trajectories from \emph{weak} to \emph{hidden strong coupling} is clearly observed. Note how the metallic shell sustains two plasmon resonances; one predominantly at the outer surface and one mostly at the inner surface. The outer resonance [green line in (a); bright green line in (b)] is highly damped because of strong radiative loss and couples only weakly to the core, due to a vanishing overlap. As such, it passes by without significantly interacting with the material resonance. In contrast, the plasmon mode that lives mostly on the inner surface (blue line), is characterized by a lower radiation loss and a better overlap with the SURMOF core. This inner plasmon mode exhibits trajectories very similar to the ones found in Figure~\ref{fig:osc_reduction}, which equivalently permits the categorization into \emph{weak}, \emph{strong}, and \emph{observable strong coupling} from the changing complex eigenfrequencies alone.

\subsection{Multiple material resonances} 
\label{sec:multipe_material_res}

\begin{figure}[t]
    \centering
    \begin{tikzpicture}[      
        every node/.style={anchor=south west,inner sep=0pt},
        x=1mm, y=1mm,
      ]   
     \node (fig1) at (0,0)
       { \includegraphics[width=1\linewidth]{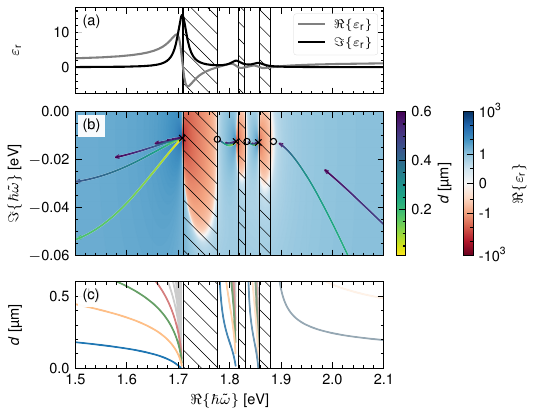}};
     \node (fig2) at (59,7)
       {\includegraphics[width=0.31\linewidth]{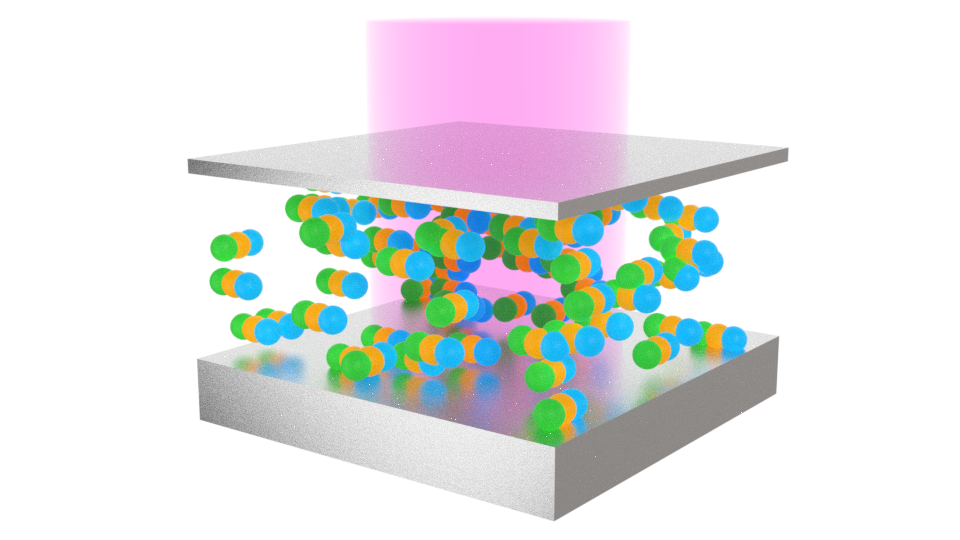}};  
    \end{tikzpicture}
    \caption{Pole trajectories: A Fabry-Perot cavity made from two silver films separated by a varying thickness $d$ is filled with a SURMOF material whose dielectric function can be described effectively by material resonances. (a) Real and imaginary part of the permittivity of the SURMOF material as extracted from quantum-chemical simulations. (b) Analogous to Figure~\ref{fig:osc_reduction}(a), the trajectories of the complex eigenfrequencies of the RSs as a function of cavity thickness $d$ (color encodes $d$). The background color corresponds to the real part of the SURMOF permittivity. The three material resonances (marked by x) generate one accumulation point each. With increasing cavity thickness, the modes dominated by the higher frequency material resonances leave the accumulation points, moving left towards the complex zeros of $\epsr$ (open circles) created by the superposition of neighboring material resonances. (c) Projection of the pole trajectories onto the real frequency axis: The \RSs/ are colored distinctly according to the Fabry-Perot mode they belong to. The dashed area marks the real frequency interval in which $\Re\{\epsr\}<0$.}
    \label{fig:three_pole}
\end{figure}

Most systems of practical relevance contain materials that support more than one material resonance. When their resonance frequencies are well separated, their coupling to the cavity mode can be investigated individually. However, when the material resonances are located in close vicinity, separating their coupling behavior becomes non-trivial.
Let us investigate the trajectories of the \RSs/ of a planar cavity filled with a realistic 3-pole model of the SURMOF material [the dielectric function is shown in Figure~\ref{fig:three_pole}(a)].
In this particular example, we do not adjust the coupling strength, but rather consider a single realistic material model: The electric dipolar polarizability of the unit cell of the SURMOF was quantum-chemically computed using the time-dependent density-functional theory. Subsequently, we applied a homogenization procedure to retrieve the dielectric function. Please see the SI for more details.   
The specific material parameters and geometry are again given in Table~\ref{tab:params}. Figures~\ref{fig:three_pole}(b) and (c) show, similarly to the previous figures, the eigenfrequencies of the \RSs/ of the hybrid system in the complex plane (b) and the projection onto the real line (c) parametrized by the cavity thickness. Similar to the case of the single material resonance discussed above, the cavity mode is split into a lower and an upper polariton branch, which emerge from the lowest-energy material pole for a vanishing cavity thickness and approach the highest-energy material zero for the largest cavity thickness, respectively. 

However, additional polariton branches emerge in-between the material resonances. These traverse the complex plane, emerging from different material resonances and approaching material zeros, but they never enter the region between the pole and the neighboring higher-energy zero of any single material resonance. This behavior leads to energy gaps in the modal dispersion. Typically, these are associated with the medium inside the FP cavity turning effectively metallic ($\Re\{\epsr(\Re\{\omega \})\}<0$; hatching in Figure~\ref{fig:three_pole}) on parts of the real-frequency axis, thereby not supporting a FP mode \cite{cao_strong_2021}. Note how the ``keep out'' regions are actually slightly larger, because the permittivity is now evaluated at complex frequencies.

\begin{figure}[t]
    \centering
    \includegraphics[width=1\linewidth]{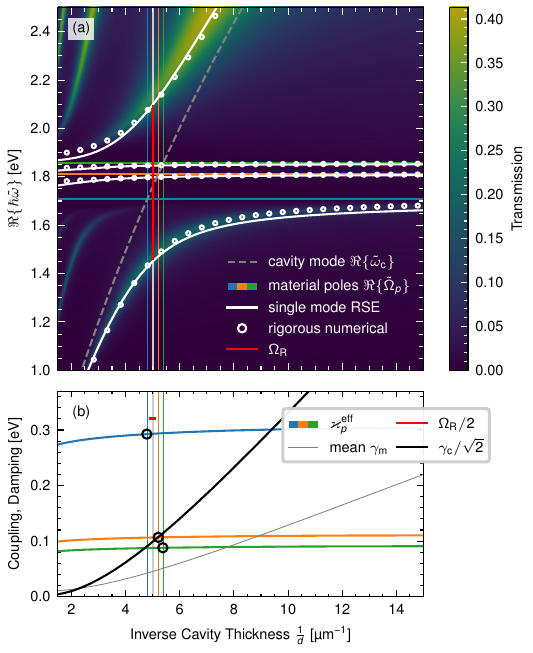}
    \caption{(a) Starting from a single \RS/ (the fundamental FP cavity mode - dashed line) and the three material resonances (solid colored lines), the four resonance frequencies of of the hybrid resonant states are predicted by our resonant state expansion (solid white lines). Circles show reference \RSs/ obtained by directly solving the nonlinear eigenproblem numerically. The transmission of the cavity shown in the background is obtained by conventional scattering simulations at real-valued frequencies. The red vertical line indicates the minimum splitting between the upper and lower polariton branches, which contains contributions from all three material resonances. (b) The effective coupling rates $\varkappa^\mathrm{eff}_p = \sqrt{|g_p \Omp|}$ (colored lines) computed from \Eq{eq:overlap} are compared to the average damping rate of all hybrid modes $\tilde \omega_m = \omega_m -\ii\gamma_m$ (solid black line) and the scaled bare cavity damping $\gamma_\mathrm{c}/\sqrt{2}$ (black line). The short red marker indicates the splitting extracted in (a).}
    \label{fig:sma_rse}
\end{figure}

Despite having access to the \RSs/ of the coupled system, an interesting challenge remains: How can one investigate the individual coupling of a cavity mode to one specific material resonance in the presence of additional material resonances?

The trajectories shown in Figure~\ref{fig:three_pole}(b), generally exhibit a behavior that is similar to the previously considered example, when strong coupling is present. Due to the multiple material resonances, the behavior is richer in its details. The hybrid \RSs/ once again swap their position. As before, we find the \RS/, that starts as the fundamental cavity mode and gets captured by the peripheral material zero, and the \RS/, that emerges from the lowest energy material resonance and asymptotically approaches the bare cavity mode. In addition we observe the pairwise exchange from higher-energy material poles (resonances) to material zeros. We conclude, that all of the material resonances are strongly coupled to the cavity mode, as none of the material poles is simply passed without being involved in these swaps.

However, evaluating the coupling strength beyond the weak coupling regime is more delicate. Simply considering the overall splitting (in red in Figure~\ref{fig:sma_rse}) provides no information on the individual coupling of the different material resonances.

To address this question, we derive an effective Hamiltonian $\hat{\mathcal{H}}$ that accounts for the hybridization between a single \RS/ of the ``empty'' photonic resonator and an arbitrary number of material resonances (refer to Section~\ref{sec:SI_RSE} of the SI for details). Then, the hybrid light-matter \RSs/ of the coupled system can be found as solutions to the eigenvalue problem:
\begin{equation}
\label{rse-eigenproblem}
    \hat{\mathcal{H}}\veccoeffrse = \tilde{\omega}\veccoeffrse\, .
\end{equation}
Here, $\tilde{\omega}$ are the hybrid eigenfrequencies, and the eigenvectors $\veccoeffrse$ contain the admixture coefficients of the original \RS/ of the empty photonic resonator (first element) and the induced polarization of the individual material resonances (subsequent elements).
For $P$ material resonances supported by the material and embedded in the resonator, $\hat{\mathcal{H}}$ has the form
\begin{equation}
    \hat{\mathcal{H}} = \begin{bmatrix}
        \tilde{\omega}_\cav+\sum_{p=1}^Pg_p&\varkappa_1&\varkappa_2&\cdots\\
        \varkappa_1&\tilde{\Omega}_1&0&\cdots\\
        \varkappa_2&0&\tilde{\Omega}_2&\cdots\\
        \vdots&\vdots&\vdots&\ddots
    \end{bmatrix}\, ,
\end{equation}
where $\varkappa_p = \sqrt{g_p \tilde \Omega_p}$, and $g_p$ is the overlap between the optical mode and the $p^{\text{th}}$ material residue:
\begin{equation}
\label{eq:overlap}
    g_p = \int_V\mb{E}_\cav(\mb{r})\cdot\varepsilon_p(\mb{r})\mb{E}_\cav(\mb{r})\text{d}\mb{r}\, .
\end{equation}

The eigenproblem in \Eq{rse-eigenproblem} exhibits great similarity to the coupled-oscillator model generalized to multiple material resonances. However, we obtain a correction to the eigenfrequency of the bare cavity mode $\omcav$ given by $\sum_p g_p$, which is absent in the phenomenological models.

 % Their precise makeup and other aspects are treated in detail in SI~\ref{sec:SI_RSE}. A peculiar consequence of the divergence of $\epsr(\omega)$ close to material poles, is the emergence of accumulation points of QNMs, as we discuss in section~\ref{sec:weak_strong_coupling}.

Let us now demonstrate the applicability of our model to the example system previously introduced in Figure~\ref{fig:three_pole}. A colormap of the observable transmission as a function of the inverse cavity thickness and the real frequency is shown in Figure~\ref{fig:sma_rse}. It is overlayed with resonance frequencies that were calculated in two ways: Given only the bare cavity \RS/ [eigenfrequency (dashed line), normalized fields] and the material resonances [eigenfrequencies (colored lines), distribution in space $\varepsilon_p(\mb{r})$], the linear eigenproblem in \Eq{rse-eigenproblem} predicts the hybrid eigenfrequencies (solid white lines). We compare these to numerical reference solutions (open white circles), obtained by directly solving the nonlinear eigenproblem posed by the cavity including all material resonances. Excellent agreement between the resonance frequencies in both of these approaches is found. As such we conclude, that the introduced resonant state expansion faithfully models the coupling between a photonic mode and multiple material resonances

Besides the accurate prediction of the hybrid eigenfrequencies, the method allows us to directly evaluate the overlaps $g_p$, giving us access to the coupling terms $\hat{\varkappa}_p$ and $\varkappa_p$ without any fitting procedures. Let us now use these to investigate whether the strong coupling of individual material resonances is observable.
In the presented example, $g_p\Omp$ is almost fully real. As such, we can restrict our analysis to individual effective retardation-free coupling rates
\begin{equation}
    \label{eq:effective_coupling}
    \varkappa^\mathrm{eff}_p = \sqrt{|g_p \Omp|}\, ,
\end{equation} 
in analogy to the coupled-oscillator model. The three $\varkappa^\mathrm{eff}_p$ from the current example are shown in Figure~\ref{fig:sma_rse} (b).

The material resonance with the lowest frequency contributes significantly more to the overall splitting [red lines in both Figure~\ref{fig:sma_rse}(a) and (b)] than the remaining two resonances. Only the coupling of the dominant material resonance exceeds the damping of all bare resonances (light and matter). In particular, the coupling rates $\varkappa^\mathrm{eff}_\mathrm{2,3}$ with the weaker material poles lie below $\gamma_\cav$ in the relevant thickness range. The coupling rates of all three material resonances exceed the average damping of the four split polariton branches (gray line in Figure~\ref{fig:sma_rse}(b)). Considering the first form of the observable strong coupling criterion [Equation~\eqref{eq:observable_strong_coupling_rabi}], one might be mislead to read this as an indication that all material resonances exhibit observable strong coupling. However, considering the average damping of all bare resonances is not a robust indicator insofar as it incorporates all far-detuned and also very-weakly-coupled material resonances, playing a vanishing role in the system dynamics. Whether these irrelevant material resonances on average have high or low damping should play no role in the investigation of the coupling regime of a distinct material resonance.

On the other hand, \emph{individually} comparing the effective coupling strength $\varkappa^\mathrm{eff}_p$ to the damping rates of the uncoupled optical mode and the $p^\text{th}$ material resonance provides a natural and robust way to generalize the second criterion for observable strong coupling [Equation~\eqref{eq:observable_strong_coupling_g}] to multiple material resonances:
\begin{equation}
    \varkappa^\mathrm{eff}_p > \sqrt{\frac{\gamma_\cav^2 +\gamma_p^2}{2}} \, .
\end{equation}
Considering the low damping of the material resonances, \textit{i.e.}, $\gamma_p \ll \gamma_\cav$, we introduce the approximation
\begin{equation}\sqrt{\frac{\gamma_\cav^2 +\gamma_p^2}{2}} \approx \frac{\gamma_\cav}{\sqrt{2}} \, ,
\end{equation}
which is shown as the black line in Figure~\ref{fig:sma_rse}(b). If the material resonances had stronger damping we could simply proceed without the approximation, which would, however, clutter Figure~\ref{fig:sma_rse}(b). The simplified criterion now reads as
\begin{equation}
    \varkappa^\mathrm{eff}_p(d) \gtrapprox \frac{\gamma_\cav(d)}{\sqrt{2}} \, .
\end{equation}
Here, we reintroduce the explicit dependence on the system parameter $\rho$, which in this case corresponds to the cavity thickness $d$, previously omitted for readability. Due to the dispersive silver mirrors, $\gamma_\cav(d)$ changes significantly with $d$. Therefore, it is relevant to compare the coupling and damping for an appropriate $d$. Here, we chose the cavity thickness $d_p$ such that the real parts of the bare cavity mode and $p^\text{th}$ material resonance match (indicated by the colored vertical lines). The black circles in Figure~\ref{fig:sma_rse}(b) indicate the different $\varkappa^\mathrm{eff}_p(d_p)$, which have to be compared to $\gamma_\cav(d_p)/\sqrt{2}$ at the same thickness.
From the generalized criterion above, we can conclude that the dominant material resonance (blue) exhibits observable strong coupling, while the second material resonance (orange) is on the verge of observable strong coupling, and for the third material resonance (green), the strong coupling is hidden by the damping of the optical cavity $\gamma_\cav$.

In summary, the presented framework enables the nuanced analysis of arbitrary combinations of material resonances with different couplings to a joint optical mode. While these calculations rely on overlap integrals of adequately normalized \RSs/, the individual coupling coefficients are phenomenologically accessible directly from the eigenfrequencies of the coupled system. A detailed explanation is provided in Section~\ref{si:inv_eigenproblem} of the SI.

% \begin{itemize}
%     \item Discuss 3 pole cavity [Figure \ref{fig:three_pole}]. Introduce Inverse Eigenproblem. (Assume the 1 Optical 3 Material Hamiltonian (denoted as the 2Nx2N Hamiltonian in {\tiny https://journals.aps.org/prb/pdf/10.1103/PhysRevB.103.L241407}): The cavity modes independently couple to the material resonances. No mixing between the different cavity modes due to the material resonances). See RSE derivation in SI. Introduce Correction.
%     \item Solve for coupling coefficients and bare optical mode frequency (with the correction). See SI for the derivation why/how the inverse eigenproblem can be solved. Shows that in this cavity we have observable strong coupling with respect to all three material resonances.
%     \item Can be used to analyze situations where one might only have strong coupling with some of the resonances [TODO: expand on this]
% \end{itemize}

% \textbf{Further references}: 
% \begin{itemize}
%     \item $\hat g = g^*$: \cite{kharel_multimode_2022}
%     \item Discussing the most general ($N_o+N_x \times N_o+N_x$) Hamiltonian: \cite{mandal_microscopic_2023}
% \end{itemize}

%\lipsum[1-2]

\section{Conclusion}
\label{sec:conclusions}
% \begin{itemize}
%     \item QNMs are well suited to investigate coupling.
%     \item three regimes
%     \item realistic material resonances (multiple Lorentz)
%     \item Outlook: Consider that typical exciton polariton already incorporates multiple "local" material resonances. Inhomogeneous broadening and/or closely spaced electronic transitions. Which material resonances are considered as separate, which as one?
%     \item Generalize to multiple optical modes. -> Points to next paper
% \end{itemize}

In this work, we have shown that photonic \RSs/ provide a rigorous and unambiguous framework for classifying regimes of light-matter interactions in open resonant systems. By following the trajectories of resonant states in the complex frequency plane as a function of some system parameter, the resonant states reveal clear signatures of weak coupling, hidden strong coupling, and spectrally resolvable (observable) strong coupling. Importantly, this perspective captures subtle effects often obscured in real-frequency observables, such as Purcell- and Lamb-like shifts, or the passage through exceptional points. The resonant states allow for a direct extraction of coupling rates, even in the presence of multiple material resonances. While for simple spherical and planar systems, the resonant states of the uncoupled (i.e., dispersionless) cavity can be obtained from analytical transcendental equations~\cite{Armitage2014a, Doost2014a, Defrance2020a}, it is straightforward to retrieve them numerically in more complex systems~\cite{both_resonant_2022, lalanne_light_2018, Weiss2016a}.

Beyond the specific examples studied here, our study highlights the importance of considering the true eigenmodes of open photonic systems for designing and interpreting light-matter interactions in complex photonic environments. Hence, the \RS/ framework provides a foundation to understand and design resonators for a broad range of applications in photonics and beyond, ranging from polariton lasing~{\ifmanycite\cite{butov_polariton_2007, schneider_electrically_2013, stoferle_ultracompact_2010}\else\cite{butov_polariton_2007, schneider_electrically_2013}\fi} and computing~{\ifmanycite\cite{ballarini_all-optical_2013, zasedatelev_room-temperature_2019, kavokin_polariton_2022, tassan_integrated_2024}\else\cite{ballarini_all-optical_2013, zasedatelev_room-temperature_2019, kavokin_polariton_2022}\fi} to polariton chemistry~\ifmanycite\cite{hutchison_modifying_2012, thomas_tilting_2019, hertzog_strong_2019}\else\cite{hutchison_modifying_2012, hertzog_strong_2019}\fi. 

From an experimental perspective, the central challenge is that the complex eigenfrequencies are not directly measurable. In the present work, we therefore view our approach primarily as a modeling and design tool: Starting from a geometric description of the resonator and a (Drude-)Lorentz representation of the constituent materials, our framework yields the hybrid eigenfrequencies and coupling rates. Whenever the underlying model faithfully captures the experimental structure, the resonant states framework can be used to interpret the underlying modal structure of measured observables.

In contrast, discrepancies between measured and simulated spectra indicate that either the material response or the geometry is not adequately represented. In such situations, it becomes desirable to extract improved effective models, or even the complex eigenfrequencies themselves, directly from real-frequency measurements. Rational approximation techniques, such as the AAA algorithm that we here employ on simulated spectra, are, in principle, capable of reconstructing poles and residues from measured spectra. While the standard AAA algorithm is sensitive to noise~\cite{nakatsukasaApplicationsAAARational2025}, noise-robust extensions suitable for experimental data have already been suggested~\cite{zivanovicAutomaticAlgorithmAAA2023}, and further developments can be expected. These approaches work best given access to both amplitude and phase of the relevant transfer function (e.g., complex reflection or transmission coefficients). Fortunately, techniques for measuring such quantities have been described in the past and can be used for this purpose \cite{pshenay-severinExperimentalDeterminationDispersion2010}. Developing and validating such reconstruction schemes in a close collaboration of theoretical and experimental groups will be an important step towards applying our \RS/-based strong-coupling criteria directly to measurements.

A remaining limitation of the presented resonant state expansion is the restriction to a single dominant photonic mode. We are currently working towards a generalization that incorporates multiple photonic modes (and corresponding counter-rotating contributions). Such a multimode extension would broaden the scope of our approach, enabling the systematic treatment of complex cavities and metasurfaces. We plan to investigate the emergence of surface-exciton polaritons \cite{gentileOpticalFieldEnhancementSubwavelength2014} and so called Berreman modes \cite{vassantBerremanModeEpsilon2012}, which both arise when an exciton creates permittivity with a negative real part. Overall, the extension will enable the treatment of platforms that host several photonic and material resonances, in line with recent effective-Hamiltonian studies of multimode strong coupling~\cite{mandal_microscopic_2023}. We anticipate that these developments will further strengthen the role of \RSs/ as a unifying language for light-matter interaction in open, lossy, and structurally complex photonic environments.

\vfill\null

\emph{Acknowledgements} -- The authors acknowledge helpful discussions with Thomas Jebb Sturges. J.D.F. acknowledges the Karlsruhe School of Optics and Photonics (KSOP).

\emph{Funding Information} -- J.D.F. and C.R. acknowledge ﬁnancial support by the
Helmholtz Association in the framework of the innovation platform ``Solar TAP''. M.N. and C.R. acknowledge support by the KIT through the ``Virtual Materials Design'' (VIRTMAT) project. T.W. and S.G. acknowledge funding by the individual DFG project WE5815/5-1. A.C.V. acknowledges support from the Visiting
Awards for High Potentials from the University of Graz.

\emph{Conflict of Interest} --
The authors state no conflict of interest.

\emph{Code and Data Availability} --
The codes used to conduct the current study are available in the github repository, www.github.com/tfp-photonics/agsurmof.
The datasets generated and analyzed during the current study are equally available in the same repository.

\emph{Parameters of the Investigated Examples}
The materials and resonator geometries in this article are inspired by the experimental and theoretical work presented in references~\cite{haldar_guest-responsive_2020} and~\cite{zerulla_multi-scale_2022}, respectively. In those, a planar cavity is formed by depositing a film of surface-metal-organic framework (Zn-SiPc-SURMOF-2) between two thin silver layers, forming the top and bottom reflectors.  While the optical properties of silver are readily available in the literature~\cite{johnson_optical_1972}, the authors of~\cite{zerulla_multi-scale_2022} developed a multi-scale framework to obtain optical properties from quantum-chemical time-dependent density functional theory (TD-DFT) simulations, which are compatible with device-scale simulations. After obtaining molecular polarizabilities from the TD-DFT simulations, these are converted to dipolar transition matrices (T-matrices). The introduced framework then provides two distinct paths to compute the optical response of macroscopic objects. Either the T-matrices can be directly used in periodic and aperiodic multiple-scattering calculations via the transition matrix formalism. Alternatively, homogenized material parameters (such as the relative permittivity $\epsr$) can be obtained for periodic and disordered arrangements of molecules, which is the approach we follow here.

To enable evaluating the \RSs/, we further require material models that allow for analytic continuation to the complex frequency plane. Material models that are rooted in a description based on ordinary differential equations naturally fulfill this requirement. These include the above-mentioned Drude and Drude-Lorentz models. The homogenized permittivities of the SURMOF material were fit to good agreement with a three-pole Lorentz-oscillator model.

% Geometrical and material parameters of the examples investigated throughout Figures~\ref{fig:observable} to~\ref{fig:sma_rse} are provided in Table \ref{tab:params} below.

\begin{table*}[t]
\resizebox{\linewidth}{!}{%
\begin{tabular}{c|p{2.9cm}p{2.9cm}p{4.7cm}p{4cm}}
\toprule
Figure & System              & Geometry                                                                         & Material Model SURMOF& Material Model Silver\\ \midrule
2,3 &
  Planar Cavity &
  \qty{10}{nm} Silver \newline variable cavity \newline thickness $d$ \newline filled with SURMOF \newline \qty{30}{nm} Silver &

  \multirow{2}{5cm}{Lorentz model: \newline $\varepsilon_\infty=1.6$ \newline 1. Pole:\newline Oscillator Strength $\sqrt{\frac{N_1 q^2}{m_1 \varepsilon_0} f_1}=2\pi\times\qty{180.26633}{THz}$ \vspace{0.1cm} \newline Resonance \newline $\omega_{0, 1} = 2\pi \times \qty{412.93727}{THz}$ \vspace{0.1cm} \newline Damping $\Gamma_{1} = 2\pi \times \qty{5.3}{THz}$}
  
  &
  
  \multirow{2}{5cm}{Lorentz model: \newline $\varepsilon_\infty=1$ \newline  Oscillator Strength\newline (or Plasma Frequency) \newline$\sqrt{\frac{N_\mathrm{Ag} q^2}{m_\mathrm{Ag} \varepsilon_0} f_\mathrm{Ag}}=\sqrt{\omega_\mathrm{p,Ag}}=2\pi\times\qty{1867.2715}{THz}$ \vspace{0.1cm}\newline Damping \newline$\Gamma=2\pi\times\qty{10.618653}{THz}$ \vspace{0.1cm}\newline Resonance \newline $\omega_{0, \mathrm{Ag}} = 2\pi \times \qty{412.93727}{THz}$}
  
  \\ \cmidrule(r){1-3}
4      & Core Shell Sphere & variable $r_\mathrm{outer} = r_2$\newline $r_1 = r_2 - \qty{20}{nm}$ \vspace{0.7cm} &                &  \\ \cmidrule(r){1-4}
5,6 & Planar Cavity & as Figure 1 & 
  \multirow{2}{5cm}{Additional Poles:\vspace{0.1cm}\newline 2. Pole:\newline $\sqrt{\frac{N_2 q^2}{m_2 \varepsilon_0} f_2}=2\pi\times\qty{65.517166}{THz}$\newline $\omega_{0, 2} = 2\pi \times \qty{438.29307}{THz}$ \newline $\Gamma_{2} = 2\pi \times \qty{6.0}{THz}$ \vspace{0.1cm}\newline 3. Pole:\newline $\sqrt{\frac{N_3 q^2}{m_3 \varepsilon_0} f_3}=2\pi\times\qty{53.923443}{THz}$\newline $\omega_{0, 3} = 2\pi \times \qty{448.79110}{THz}$ \newline $\Gamma_{3} = 2\pi \times \qty{6.2}{THz}$} & \\
  &   \vspace{-0.2cm}\hspace*{0.1cm}\includegraphics[width=5.5cm]{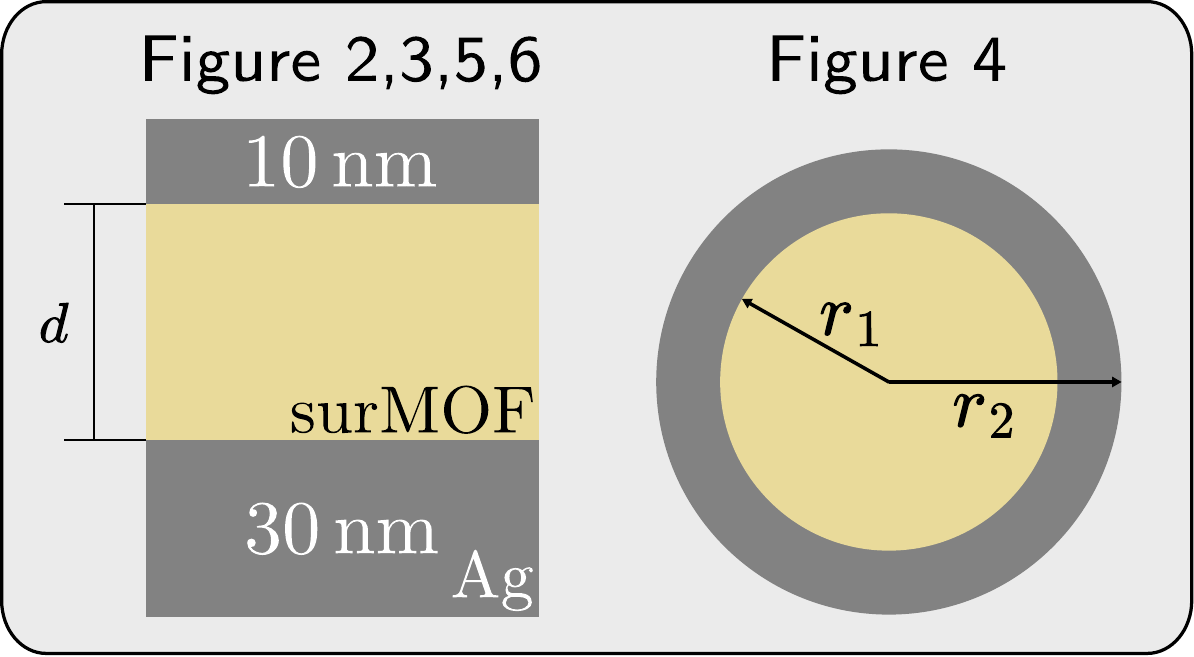} \vspace{-0.4cm} &    &  \; \newline \newline \newline \newline \newline &  \\ \bottomrule
\end{tabular}%
}
\caption{Geometry and material models of the treated examples.}
\label{tab:params}
\end{table*}

% Markus Model
% Drude (-Lorentz): \newline Oscillator Strength\newline (or Plasma Frequency) \newline$\sqrt{f_\mathrm{Ag}}=2\pi\times\qty{1867.2715}{THz}$\newline Damping \newline$\Gamma=2\pi\times\qty{10.618653}{THz}$ \newline Resonance \newline $\omega_{0, \mathrm{Ag}} = 2\pi \times \qty{412.93727}{THz}$

%Drude Lorentz: \newline 1. Pole:\newline Oscillator Strength $\sqrt{f_1}=2\pi\times\qty{180.26633}{THz}$\newline Resonance \newline $\omega_{0, 1} = 2\pi \times \qty{412.93727}{THz}$ \newline Damping $\Gamma_{1} = 2\pi \times \qty{5.3}{THz}$} & Drude: \newline Plasma Frequency \newline$\omega_\mathrm{plas, Ag}=2\pi\times\qty{2124.9348}{THz}$\newline Damping \newline$\Gamma=2\pi\times\qty{10.450943}{THz}$

\FloatBarrier

\putbib
\end{bibunit}
\clearpage

\begin{bibunit}
% --- SUPPLEMENTARY INFORMATION STARTS ---
% Reset internal flag so we can call \maketitle again
\makeatletter
\setcounter{page}{1} % optional: reset page numbering
\def\@title@maketitle@done{} % <--- This is the key line
\makeatother

\title{Supplementary information: \papertitle}
\paperauthors
\maketitle

%\section{Supplementary Information}
\markboth{\MakeUppercase{Supplementary Information}}{\MakeUppercase{Supplementary Information}}

\subsection{Translating Lorentz to pole-based material model}

\newcommand{\omhat}{\hat\omega_{0,j}}

Here, we provide the link between Eq.~\eqref{eq:lorentz_model} and Eq.~\eqref{eq:material_poles}. For that purpose, the fractions on the right-hand side of Eq.~(\ref{eq:lorentz_model}) are expanded as a sum over simple poles as
\begin{equation} \label{eq:partial_fraction}
\frac{f_j}{\omega_{0,j}^2-\omega^2 - \ii\omega\Gamma_j} = \frac{A_j}{\omega - \Om_{j,+}} + \frac{B_j}{\omega - \Om_{j,-}},
\end{equation}
where $\Om_{j,+} = \omhat-\ii\Gamma_j/2$ and $\Om_{j,-} = -\Om_{j,+}^*$.
Defining $\omhat^2\equiv\omega_{0,j}^2-\frac{1}{4}\Gamma_j^2$, we find that $B_j$ is given by

\begin{equation}
B_j = \frac{f_j}{2 \omhat} \, ,
\end{equation}
and $A_j = -B_j$.

When further adding a background permittivity $\varepsilon_\infty$ to encapsulate resonances outside the frequency domain of interest, Eq.~(\ref{eq:lorentz_model}) can be written as
\begin{equation}
\begin{aligned}
    &\epsr(\omega) 
    = \varepsilon_{\infty}+\\
    &\frac{Nq^2}{m_0\varepsilon_0}
       \sum_j \left( \frac{A_j}{\omega - \Om_{j,+}}
       + \frac{B_j}{\omega - \Om_{j,-}} \right)\, .
\end{aligned}
\end{equation}
Separating the two poles associated to each $j$, the latter expression takes the form of Eq.~(\ref{eq:material_poles}) in the main text:
\begin{equation}
        \epsr\po = \varepsilon_{\infty}-\sum_p \frac{\varepsilon_p}{\omega - \Om_{p}}\, .
\end{equation}
Here, we have introduced $\varepsilon_p $ as

\begin{equation}
    \varepsilon_p = \pm \frac{Nq^2}{m_0 \varepsilon_0} \frac{f_j}{2 \omhat}\, ,
\label{eq:sigma}
\end{equation} 
with the $\pm$ signs corresponding to the $\Om_{j,\pm}$ poles in Eq.~(\ref{eq:partial_fraction}), respectively. In the single-mode approximation we will only consider the poles $\Om_{j,+}$ with positive real part and the associated $+$ sign in \eqref{eq:sigma}. The introduced $\varepsilon_p$ are closely related to the generalized conductivity tensors $\mathbf{\sigma}$ of Reference~\cite{muljarov_resonant-state_2016} via $\ii \boldsymbol{\sigma} = - \hat{I} \varepsilon_p $. The $3\times3$ identity matrix $\hat{I}$ accounts for the conversion from general anisotropic media with tensorial material properties to the simplified case of isotropic media considered here. The factor $\ii$ difference in notation ensures that for the considered Lorentz-media the quantities introduced in the derivation below are predominantly real and positive.

\subsection{Effective Hamiltonian in the single-mode approximation}
\label{sec:SI_RSE}

We derive a first-principles theory on the hybridization of a single \RS/ with an arbitrary number of material resonances. To do that, we start with the source-free Maxwell's curl equations in the frequency domain, which can be written  in a compact notation as~\cite{weiss_how_2018}:
\begin{equation} \label{eq:max_eq}
    \left[\frac{\omega}{c}\hat{\mathbb{P}}(\mb{r};\omega)-\hat{\mathbb{D}}\right]\mathbb{F}(\mb{r}) = 0\, .
\end{equation}
Here, we have introduced the material tensor $\hat{\mathbb{P}}\carsten{(\mb{r};\omega)}$, assuming non-magnetic, achiral and isotropic materials, as:
\begin{equation}
    \hat{\mathbb{P}}(\mb{r};\omega) = 
    \begin{bmatrix}
        \epsr(\mb{r};\omega)\hat{I} &\mathbf{0}\\
        \mathbf{0}&\hat{I}
    \end{bmatrix}\, .
\end{equation}
In the above, $\hat{I}$ is the $3\times3$ identity matrix. The matrix $\hat{\mathbb{D}}$ is given by:
\begin{equation}
    \hat{\mathbb{D}} = \begin{bmatrix}
        \mathbf{0} & \nabla\times\\
        \nabla\times & \mathbf{0}
    \end{bmatrix}\, .
\end{equation}
In addition, we have arranged the electric and magnetic fields in the vector $\mathbb{F}(\mb{r})$ as
\begin{equation}
    \mathbb{F}(\mb{r}) = \begin{bmatrix}
        \mb{E}(\mb{r})\\
        \ii{}Z_0\mb{H}(\mb{r})
    \end{bmatrix}\, ,
\end{equation} 
where $Z_0$ is the impedance of vacuum.

In this context, we are interested in understanding the interaction between an ``empty'' optical cavity filled with a material characterized by some host permittivity $\varepsilon_{\cav}(\omega)$, and the resonances of a target material that would be introduced in the cavity. Upon filling the cavity with the material, the permittivity is modified by a series of $P$ additional material poles:
\begin{equation} \label{eq:mat_poles_3}
    \epsr(\mathbf{r};\omega) = \varepsilon_\cav(\mathbf{r};\omega) - \sum_{p=1}^P\frac{\varepsilon_p(\mb{r})}{\omega-\Omp}\, ,
\end{equation}

Introducing Eq.~(\ref{eq:mat_poles_3}) in Eq.~(\ref{eq:max_eq}), and moving the terms due to the new material to the right-hand side, we get:
\begin{equation}
    \left[\frac{\omega}{c}\hat{\mathbb{P}}_\cav(\mb{r};\omega)-\hat{\mathbb{D}}\right]\mathbb{F}(\mb{r}) = -\frac{\omega}{c} \sum_{p=1}^P\begin{bmatrix}
        \mb{P}_p(\mb{r})\\0
    \end{bmatrix}\, .
\end{equation}
Here, $\mathbf{P}_p(\mb{r})$ is the polarization density induced by each new material pole on the cavity. The latter obeys the equation
\begin{equation} \label{eq:pol}
    (\omega -\Omp\mb)\mb{P}_p(\mb{r}) = -\varepsilon_p\mb{E}(\mb{r})\, ,
\end{equation}
which is essentially the Fourier transform of the rate equation for the polarization density of the $p^\mathrm{th}$ material pole in the presence of an electric field.

Hence, the effect of filling the cavity with a dispersive material manifests as the emergence of polarization currents that perturb the eigenmodes of the empty cavity. The perturbed electric field can then be found with the electric part of the Green's function of the unperturbed system $\hat{\mathbf{G}}_\cav(\mb{r},\mb{r}')$, by integrating over the volume $V$ filled with the material:
\begin{equation} \label{eq:lipp}
    \mb{E}(\mb{r}) = -\frac{\omega}{c}\sum_{p=1}^P\int_{V}\hat{\mathbf{G}}_\cav(\mb{r},\mb{r}')\cdot \mb{P}_p(\mb{r}')\text{d}\mb{r}'\, .
\end{equation}

We now assume the dynamics of the empty cavity are primarily driven by a single optical mode, with normalized electric field $\mb{E}_\cav(\mb{r})$ and eigenfrequency $\tilde{\omega}_\cav$. As a result, the Green's function can be approximately written as~\cite{weiss_how_2018}:
\begin{equation} \label{eq:mittag}
    \hat{\mathbf{G}}_\cav(\mb{r},\mb{r}')\approx c\frac{\mb{E}_\cav(\mb{r})\otimes \mb{E}_\cav(\mb{r'})}{\omega - \tilde{\omega}_\cav}\, .
\end{equation}
In this case, $\otimes$ denotes the dyadic product. 
Inserting Eq.~(\ref{eq:mittag}) in Eq.~(\ref{eq:lipp}), we find that the perturbed eigenfield can be written as $\mathbf{E}(\mb{r}) = a\mb{E}_\cav(\mb{r})$, with the expansion coefficient $a$ given by:
\begin{equation} \label{eq:aO}
\begin{aligned}
    a &= -\sum_{p=1}^P\frac{\omega}{\omega - \tilde{\omega}_\cav}\int_{V}\mb{E}_\cav(\mb{r})\cdot\mb{P}_p(\mb{r})\text{d}\mb{r}\\ &= -\frac{\omega}{\omega - \tilde{\omega}_\cav}\sum_{p=1}^P n_p b_p\, .
\end{aligned}
\end{equation}
Here, we have introduced the abbreviation 
\begin{equation} \label{eq:b_p}
b_p \equiv \frac{1}{n_p} \int_{V}\mb{E}_\cav(\mb{r})\cdot\mb{P}_p(\mb{r})\text{d}\mb{r}\, .
\end{equation}
The coefficients $b_p$ can be understood as the projection of the empty cavity field onto the polarization induced by the $p^{\text{th}}$ additional material pole. Note how the unconjugated form of the projection is a direct consequence of the unconjugated field $\mb{E}_\cav(\mb{r})$ in the outgoing part of the Green's function in Eq.~\eqref{eq:mittag}. The terms $n_p$ are introduced to allow for an appropriate rescaling of the coefficients later on. Knowledge of the $a$ and $b_p$ coefficients is all that is needed to describe the degree of light-matter hybridization of the eigenmodes of the filled cavity.
%Physically, it provides a measure of how much of the induced polarization is induced by the optical mode.

However, we still require additional equations to determine the coefficients $b_p$. We first note that we can write Eq.~(\ref{eq:pol}) as:
\begin{equation} \label{eq:pol2}
    (\omega -\Omp)\mb{P}_p = -\varepsilon_p a\mb{E}_\cav(\mb{r})\, .
\end{equation}
Second, we take the scalar product of Eq.~(\ref{eq:pol2}) with $\mb{E}_\cav(\mb{r})$ on the left and integrate over $V$ to yield:
\begin{equation} \label{eq:bp}
    \omega n_p b_p = \Omp n_p b_p - g_pa\, .
\end{equation}
The coupling term $g_p$ takes the form of the overlap integral
\begin{equation}
    g_p = \int_V\mb{E}_\cav(\mb{r})\cdot\varepsilon_p\mb{E}_\cav(\mb{r})d\mb{r}\, .
\end{equation}

We can now rewrite Eq.~(\ref{eq:aO}) and Eq.~(\ref{eq:bp}) as a linear eigenvalue problem for $\omega$. For that purpose, we substitute the $\omega n_p b_p$ terms in Eq.~(\ref{eq:aO}) with the help of Eq.~(\ref{eq:bp}). After rearranging the resulting expression we get
\begin{equation}
    \omega a = \left(\tilde{\omega}_\cav+\sum_{p=1}^Pg_p\right)a - \sum_{p=1}^P\Omp n_p b_p\, .
\end{equation}

Hereon, we make the replacement $\omega \rightarrow\tilde{\omega}$ to distinguish the resulting eigenfrequencies from an arbitrary frequency. The resulting system of equations can be written in matrix form:
\begin{equation}
    \hat{\mathcal{H}}\veccoeffrse = \tilde{\omega}\veccoeffrse\, ,
\end{equation}
with the eigenvector $\veccoeffrse$ given by
\begin{equation}
    \veccoeffrse = \begin{bmatrix}
        a\\
        b_1\\
        \vdots\\
        b_P
    \end{bmatrix}\, ,
\end{equation} 
and the ``effective Hamiltonian'' $\hat{\mathcal{H}}$:
\begin{equation}
    \hat{\mathcal{H}} = \begin{bmatrix}
        \tilde{\omega}_\cav+\sum_{p=1}^Pg_p&-n_1\Om_1&-n_2\Om_2&\cdots\\
         -g_1/n_1&\Om_1&0&\cdots\\
         -g_2/n_2&0&\Om_2&\cdots\\
        \vdots&\vdots&\vdots&\ddots
    \end{bmatrix}\, .
\end{equation}
The eigenvectors $\veccoeffrse$ quantify the degree of hybridization between the optical eigenmode and the material resonances. The strength of the interaction is determined by the off-diagonal elements of $\hat{\mathcal{H}}$.

Choosing $n_p = -\sqrt{g_p/\Om_p}$ we arrive at a complex symmetric
\begin{equation}
    \hat{\mathcal{H}} = \begin{bmatrix}
        \tilde{\omega}_\cav +  \sum_{p=1}^Pg_p&\varkappa_1&\varkappa_2&\cdots\\
        \varkappa_1&\Om_1&0&\cdots\\
        \varkappa_2&0&\Om_2&\cdots\\
        \vdots&\vdots&\vdots&\ddots
    \end{bmatrix}\, ,
\end{equation}
with $\varkappa_p = \sqrt{g_p \Om_p}$ and an appropriate branch choice for the square root.

% The following derivation differs in units from the typical resonant state expansions. Here we use SI units, consistent with the main text, while RSE is commonly expressed in cgs units.

% \begin{equation}
% \begin{aligned}
%     \left[\omega \underbrace{\begin{bmatrix}
% \varepsilon_0\epsr(\pos, \omega) & \\
%  & \mu_0
% \end{bmatrix}}_{\bP \po}  - \underbrace{\begin{pmatrix}
%  & \nabla\times\\
% \nabla\times & 
% \end{pmatrix}}_{\bD} \right] &  \underbrace{\begin{bmatrix}
% \fE \po \\ \ii \fH \po
% \end{bmatrix}}_{\bF\po} \\=& \underbrace{\begin{bmatrix}
% -\ii\fJ\po \\ 0
% \end{bmatrix}}_{\bJ\po}
% \end{aligned}
% \end{equation}

% compactly written as 
% \begin{equation}
%     \underbrace{[\omega \bP \po - \bD]}_{\bM \po} \bF\po = \bJ \po
% \end{equation}
% Defining the maxwell operator $\bM$.

\subsection{Inverse Eigenproblem}
\label{si:inv_eigenproblem}

\begin{figure}[t]
    \centering
    \includegraphics[width=1\linewidth]{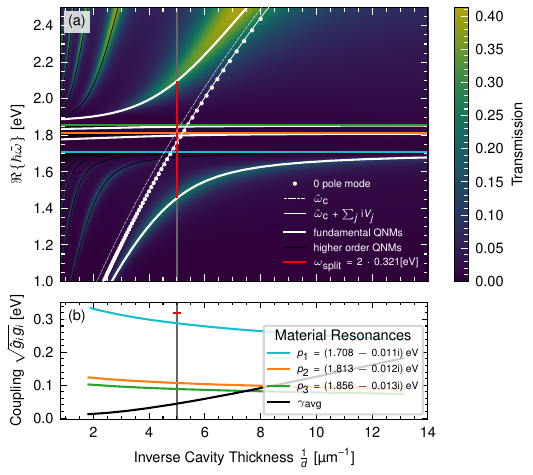}
    \caption{Extracting the coupling coefficients from the complex \RS/ frequencies: (a) We only consider \RSs/ associated to the fundamental Fabry-Perot resonance. The cavity thickness at which the separation between the upper- and lowermost \RS/ is minimized is indicated by the red line. This minimum separation corresponds to the Rabi frequency typically extracted from experiments. (b) Coupling rates between the optical and material poles, which was extracted from the \RS/ trajectories by solving the inverse eigenproblem posed by Eq.~(\ref{eq:big_hamiltonian}). The short red marker indicates the Rabi frequency extracted in (a).}
    \label{fig:inverse_eigenproblem}
\end{figure}

Here, we show how the coupling rates can be found phenomenologically in the presence of more than one material resonance if the eigenfrequencies of the coupled system are known.
Let us start from the effective Hamiltonian obtained above (let us exemplarily use 3 material resonances):
\begin{equation}
\begin{bmatrix}
\dotcoeff{\cav} \\
\dotcoeff{1} \\
\dotcoeff{2} \\
\dotcoeff{3}
\end{bmatrix}=
\underbrace{
\begin{bmatrix}
\om c + K & \coupling_1 & \coupling_2 & \coupling_3\\
\coupling_1 & \Om_1 & 0 & 0\\
\coupling_2 & 0 & \Om_2 & 0\\
\coupling_3 & 0 & 0 & \Om_3 
\end{bmatrix}}_{\mathcal{H}_3}
\begin{bmatrix}
\coeff{\cav} \\
\coeff{1} \\
\coeff{2} \\
\coeff{3}
\end{bmatrix}\, .
\label{eq:big_hamiltonian}
\end{equation}
Here, we have summarized the correction to the eigenfrequency of the cavity mode as $K$.
$\mathcal{H}$ must have eigenvalues corresponding to the resonance frequencies of the coupled system, which can be found directly by solving the nonlinear eigenproblem posed by the sourceless Maxwell equations with outgoing boundary conditions. As the characteristic polynomial contains only contains $\hggi$, rather than $\coupling_p$ we consider $\hggi$ as the free variable. From the four known eigenvalues, we can construct a system of four equations ({\it e.g.}, for the coefficients in the polynomial). As a consequence, we can unambiguously solve for the four unknowns $\hggi$ and $\tilde \omega_\mathrm{c}$. 

Let us write this out step by step. The characteristic polynomial is defined as:
\begin{equation}
C(\lambda) = 
\begin{vmatrix}
\om c + K - \lambda & \coupling_1 & \coupling_2 & \coupling_3\\
\coupling_1 & \Om_1 - \lambda & 0& 0\\
\coupling_2 & 0 & \Om_2 - \lambda & 0\\
\coupling_3 & 0 & 0 & \Om_3 - \lambda
\end{vmatrix}\, .
\end{equation}
Its zeros correspond to the eigenvalues of the Hamiltonian, which in turn correspond to the \RS/ frequencies, which can be derived analytically or numerically. We use the second row to develop the determinant as:
\begin{equation}
%\normalsize
\begin{aligned}
    C(\lambda) &= (\Om_1 - \lambda) \underbrace{\begin{vmatrix}
\om c + K - \lambda  & \coupling_2 & \coupling_3\\
\coupling_2 & \Om_2 - \lambda & 0\\
\coupling_3 & 0 & \Om_3 - \lambda
\end{vmatrix}}_{C_{\backslash{}\Om_1}(\lambda)} \\& \hspace{1cm} - \coupling_1 \begin{vmatrix}
\coupling_1  & \coupling_2 & \coupling_3\\
0 & \Om_2 - \lambda & 0\\
0 & 0 & \Om_3 - \lambda
\end{vmatrix} \\&= (\Om_1 - \lambda) C_{\backslash{}\Om_1}(\lambda) - \coupling_1^2 \begin{vmatrix}
 \Om_2 - \lambda & 0\\
0 & \Om_3 - \lambda
\end{vmatrix} \\ &= (\Om_1 - \lambda) C_{\backslash{}\Om_1}(\lambda) - \coupling_1^2 (\om{x,2} - \lambda)(\om{x,3} - \lambda)\, .
\end{aligned}
\end{equation}
Here, $C_{\backslash{}\Om_1}(\lambda)$ is the characteristic polynomial for the system after eliminating $\Om_1$. Resolving this recursion, we can write down the characteristic polynomial for an arbitrary number of material resonances explicitly as:
\begin{equation}
\begin{split}
    C(\lambda) =& \underbrace{\left[\prod_p (\Om_p - \lambda)\right]}_{\equiv R(\lambda)} (\om c + K - \lambda) \\
    &- \sum_p \hggi \underbrace{\prod_{j\backslash{}p}(\Om_j-\lambda)}_{\equiv Q_p(\lambda)}\, .
\end{split}
\end{equation}
From the definitions of $R(\lambda)$ and $Q_p(\lambda)$, we can derive:
\begin{equation}
    Q_i(\lambda) = R(\lambda)/(\om{x,i}-\lambda)\, .
\end{equation}
We can now use these expressions to write down the system of equations resulting from the different eigenvalues $\lambda_l$ compactly as
\begin{equation}
    C(\lambda_l) = R(\lambda_l)\left[\om{o} - \lambda_l - \sum_i \frac{\hggi}{\om{x,i}-\lambda_l} \right] = 0\, .
\end{equation}
When knowing $\lambda_l$ and by using $R(\lambda_l) \neq 0$, the remaining system of equations is easily solved.

The results are shown in Figure~\ref{fig:inverse_eigenproblem}. A python implementation to solve the inverse eigenproblem is provided in the accompanying repository \cite{Fischbach_Source_Code_and_2025}.

% \subsection{Essential Singularities of the Scattering Response close to Material Poles}
% \label{si:accumulation_point}

% Higher order optical modes are progressively more detuned from the material resonances. As a result, one of their replicas will lie asymptotically (with the order of the cavity mode) closer to the material resonance. As the uncoupled cavity supports a (countable) infinite number of resonances at ever increasing frequencies, an infinite number of replicas will emerge close to the material resonance, forming an accumulation point of poles of the Green's dyadic. Such accumulation points are visible in Figures~\ref{fig:osc_reduction},\ref{fig:core_shell} and \ref{fig:three_pole}. 

% The emergence of such accumulation points can alternatively be understood by considering the diverging permittivity at the material poles. In particular, in an environment $\Theta_\delta = \{|z - \Omp| < \delta: z\in \mathbb{C}\}$ around the pole $\Omp$ there exists a value $K(\delta)$ such that every value $\epsilon$ with $|\epsilon| > K(\delta)$ is assumed within the environment, i.e. $\varepsilon(z_\epsilon)=\epsilon, z_\epsilon \in \Theta_\epsilon$. 

\putbib
\end{bibunit}

\end{document}